\documentclass[preprint,aps,floats,nofootinbib,amssymb,superscriptaddress]{revtex4}
\pdfoutput=1
\usepackage[sort&compress]{natbib}
\usepackage{epsf,epsfig}
\usepackage{amsmath}
\usepackage{subfigure}
\usepackage{graphicx}
\usepackage{color}
\usepackage{mathrsfs}

\newcommand{\be}{\begin{equation}}
\newcommand{\ee}{\end{equation}}
\newcommand{\ba}{\begin{array}}
\newcommand{\ea}{\end{array}}
\newcommand{\bea}{\begin{eqnarray}}
\newcommand{\eea}{\end{eqnarray}}
\newcommand{\no}{\nonumber}
\newcommand{\nn}{\nonumber}

\newcommand{\bi}{\begin{itemize}}
\newcommand{\ei}{\end{itemize}}
\newcommand{\bal}{\begin{aligned}}
\newcommand{\eal}{\end{aligned}}

\newcommand{\wplus}{W^{+}}
\newcommand{\wminus}{W^{-}}
\newcommand{\edote}[2]{ (\epsilon_{#1} \cdot \epsilon_{#2} ) }
\newcommand{\edotqmp}[3]{ (\epsilon_{#1} \cdot (q_{#2} -p_{#3})) }
\newcommand{\pdote}[2]{ (p_{#1} \cdot \epsilon_{#2} ) }

\newcommand{\eeee}[4]{ (\epsilon_{#1} \cdot \epsilon_{#2} ) (\epsilon_{#3} \cdot \epsilon_{#4}) }
\newcommand{\eepepe}[6]{ (\epsilon_{#1} \cdot \epsilon_{#2} ) (p_{#3} \cdot \epsilon_{#4} )(p_{#5} \cdot \epsilon_{#6}) }
\newcommand{\fracp}[2]{ \left( \frac{#1}{#2} \right)}
\newcommand{\lrp}[1]{ \left( #1 \right) }

\makeatletter

\begin{document}

\preprint{ICCUB-14-048}

\title{\vspace*{1.in} Unitarity and causality constraints in composite Higgs models}

\author{Dom\`enec Espriu}\affiliation{Departament d'Estructura i Constituents de la Mat\`eria,
Institut de Ci\`encies del Cosmos (ICCUB), \\
Universitat de Barcelona, Mart\'i Franqu\`es 1, 08028 Barcelona, Spain}
\author{Federico Mescia}\affiliation{Departament d'Estructura i Constituents de la Mat\`eria,
Institut de Ci\`encies del Cosmos (ICCUB), \\
Universitat de Barcelona, Mart\'i Franqu\`es 1, 08028 Barcelona, Spain}

\vspace*{2cm}

\thispagestyle{empty}

\begin{abstract}
We study the  scattering of longitudinally
polarized $W$ bosons in extensions of the
Standard Model (SM) where anomalous Higgs couplings to gauge sector and  
higher-order ${\cal O}(p^4)$ 
operators are considered. These new couplings with respect to the 
Standard Model should be thought 
of as the low energy remnants of some  
new dynamics involving the electroweak symmetry breaking sector. 
By imposing unitarity and causality constraints 
on the $WW$ scattering amplitudes we find relevant restrictions on the possible  
values of the new couplings and the presence of new dynamical resonances above $300$ GeV. 
We investigate the properties of these new resonances and their experimental 
detectability. Custodial symmetry is assumed to be exact throughout,
and the calculation avoids using the Equivalence Theorem as much as possible.
\end{abstract}

\maketitle

\section{Introduction}
Providing tools to assess the nature of the Higgs-like boson discovered at the LHC~\cite{atlas,cms} 
is probably the most urgent task that theorists face in our time. New runs will in due time clarify whether
the Higgs particle is truly elementary or there is a new scale of compositeness associated to it. In the
latter case there should be a new strongly interacting sector, an extension of the Standard Model (SM) that
conventionally is termed the extended electroweak symmetry breaking sector (EWSBS). All evidence suggests that
the scale possibly associated to the EWSBS may be substantially larger than the electroweak scale
$v=246$ GeV, but it should not go beyond a few TeV. Otherwise the mass of its lightest
scalar resonance becomes unnatural and very difficult to sustain~\cite{scaleofcompositeness}.

Of course the Higgs could be elementary and  the Minimal Standard Model (MSM) realized in nature but then
some fundamental questions of elementary particle physics would remain unanswered: there would be no natural
dark matter candidate ---not even an axion, no hope of understanding the flavour puzzle and perhaps even the 
vacuum of the theory be unstable and jeopardize our whole picture of the universe (see \cite{degrassietal}
for updated results).

Effective Lagrangians of Higgs and gauge bosons have already extensively used to study current 
LHC data~\cite{conchaetal,Concha} combining also in some case LEP and flavour data. 
This approach has the advantage to be model-independent
but the drawback is that number of operators is usually large and the choice
of a convenient basis is subject of intense debate \cite{debate}. 
Here  we are only interested in $WW$ scattering and 
work in the custodial limit. Therefore, only a restrict number of operators have to be considered.
The effective Lagrangian is
\bea
\label{eq:1}
\mathcal{L} & = &  - \frac{1}{2} {\rm Tr} W_{\mu\nu} W^{\mu\nu} - \frac{1}{4} {\rm Tr} B_{\mu\nu} B^{\mu\nu} 
+ \frac{1}{2} \partial_{\mu} h \partial^{\mu} h  - 
\frac{M_H^2}{2} h^{2} - d_{3} (\lambda v)  h^{3} -  d_{4} \dfrac{\lambda}{4} h^{4} \\ \nn 
& & + \frac{v^{2}}{4} \lrp{1+2 a\fracp{h}{v}+ b \fracp{h}{v}^{2}+...} {\rm Tr}\, D_{\mu}U^{\dagger}D^{\mu}U 
+ \sum_{i=0}^{13} a_{i} \mathcal{O}_i\,.
\eea
where 
\be
\label{eq:2}
U = \exp \lrp{i~\frac{w \cdot \tau}{v} }\quad\text{and,}\quad D_{\mu} U =  \partial_{\mu}U + 
\frac{1}{2} i g W_{\mu}^{i} \tau^{i} U - \frac{1}{2} i g' B_{\mu}^{i} U \tau^{3}.
\ee
Here the $w$ are the three Goldstone of the global group $SU(2)_L\times SU(2)_R\to SU(2)_V$. This symmetry breaking 
is the minimal pattern to  provide the longitudinal components to the $W^\pm$ and $Z$ and emerging from phenomenology. 
Here, the Higgs field $h$ is a gauge and $SU(2)_L \times SU(2)_R$ singlet. 
Larger symmetry group could be adopted~\cite{composite} and consequently further Goldstone bosons may exist ---the Higgs might  be 
one of them. But all them eventually should acquire masses, drop from an extended unitary matrix $U$ and could 
be parameterized by a polynomial expansion. 
The operators $\mathcal{O}_{i}$ include the complete set of   ${\cal O}(p^4)$ operators defined 
in~\cite{ECHL,Espriu:2012ih}. Of these only two $O_4$
and $O_5$ will contribute to $W_LW_L$ scattering in the custodial limit:
\be
\mathcal{O}_{4} = {\rm Tr}\left[ V_{\mu}V_{\nu} \right]{\rm Tr}\left[ V^{\mu}V^{\nu} \right] 
\qquad
\mathcal{O}_{5} = {\rm Tr}\left[ V_{\mu}V^{\mu} \right]{\rm Tr}\left[ V_{\nu}V^{\nu} \right],
\ee
where $V_{\mu} = \left( D_{\mu} U \right) U^{\dagger}$. When writing eq.~(\ref{eq:1}) we have assumed
the well-established chiral counting rules to limit the number of operators to the ${\cal O}(p^4)$ ones.

The parameters $a$ and $b$ control the coupling of the Higgs to the gauge sector~\cite{composite}. 
Couplings containing higher powers of $h/v$ do not enter $WW$ scattering and they have not
been included in (\ref{eq:1}). We have also introduced two additional parameters
 $d_{3}$, and $d_{4}$ that parameterize the three- 
and four-point interactions of the Higgs field\footnote{We bear in mind that this is not the most 
general form of the Higgs potential and in fact additional counter-terms are needed beyond the 
Standard Model\cite{dobado}
but this does not affect the subsequent discussion for $W_LW_L$ scattering}. 
The MSM  case corresponds to setting $a=b=d_{3}=d_{4}=1$ in Eq. (\ref{eq:1}). 
Current LHC results give the following bounds for $a$, $a_{4,5}$:
\be
\label{eq:bound_a_a4_a5}
a= [0.67,1.33],  \qquad  a_4= [-0.094,0.10], \qquad a_5= [-0.23,0.26]\qquad 
90\%\text{CL}
\ee 
see~\cite{Falkowski:2013dza,Concha}~\footnote{
Our   $a$ and $a_{4,5}$ coefficients stand for $a=1-\xi c_H/2$, $a_4=\xi^2 c_{11}$ and 
$a_5=\xi c_{6}$ of ref.~\cite{Concha}. $c_H$ range comes from the values of Set A in table 4 and $c_{6,11}$ 
are from table 8 of ref.~\cite{Concha}.}.
The present data clearly favours values of $a$ close to the MSM value, while the $a_4$ and
$a_5$ are still largely unbounded. The parameter $b$ is almost totally
undetermined at present. Other very important parameters are $a_1$, $a_2$ and $a_3$, entering the oblique and triple
gauge coupling. Bounds on the oblique corrections are quite constraining~\cite{oblique}, while the 
triple electroweak gauge coupling has already been measured with a level of precision\cite{triple} similar to LEP. 
Some results on the $\gamma\gamma W^+W^-$ coupling are also available\cite{gammagamma}.

When $a$ and $b$ depart from their MSM values $a=b=1$  the theory becomes 
unrenormalizable in the conventional sense, 
although at the one-loop level $W_LW_L$ scattering can be rendered
finite by a suitable redefinition of the coefficients $a_4$ and $a_5$ and $a$ (together with $v$, $H_H$ and $\lambda$). 
The relevant counter-terms have been worked out in~\cite{EMY,dobado} using 
the Equivalence Theorem~\cite{ET,esma} (i.e. replacing longitudinally polarized $W_L$ and $Z_L$ by the corresponding 
Goldstone bosons $w$ and $z$). This approximation is appropriate to obtain the relevant counter-terms for
$W_LW_L$ scattering and in~\cite{herreroetal} the renormalization is being extended to the remaining $a_i$ 
counter-terms ($i\ne4,5$).

In this  work we extend the previous analysis~\cite{Espriu:2012ih} of unitarized $W_LW_L$ scattering 
to the case $a\ne 1$ and $b\ne1$, namely anomalous Higgs couplings to the gauge sector are now considered. 
More specifically, we will vary $a$, $b$ as well as the $a_{4,5}$ parameters  within the experimental bounds  
of eq.~(\ref{eq:bound_a_a4_a5}).
We use the Inverse Amplitude Method (IAM)~\cite{iam} to enforce the unitarity of longitudinally polarized $WW$ amplitudes 
up to the ${\cal O}(p^4)$. The calculation of the amplitude is done  avoiding the
use of the Equivalence Theorem as much as possible. The reason for this is that at the relatively low
energies we are considering, the replacement of the $W_L$ and $Z_L$ by $w$ and $z$ is problematic in order
to make accurate predictions. In the next sections, we will give examples of how misleading the ET can be if the
right kinematical conditions are not met. 

As in the previous work~\cite{Espriu:2012ih}, we found that new dynamical 
resonances can appear in the parameter space of $a_{4,5}$ 
for given values of $a$ and $b$ even though for values of $a>1$ the allowed region 
is drastically reduced by the causality constraint.
More specifically, for $a\le1$ and $b$ free, the overall picture is very similar to one 
in~\cite{Espriu:2012ih} where the case $a=b=1$  (experimentally favoured so far) was studied. In the scalar 
channel for example,  new dynamical resonances go from masses as low as 300 GeV to 
nearly as high as the cutoff of the method 
of $4\pi v\simeq 3$~TeV, with rather narrow widths typically from $10$ to $100$~GeV. In the vector channel the
lowest achievable masses range from about 600 GeV up to the cutoff, with widths going from 5  to about 50 GeV. 
For  $a>1$, the picture is drastically different with respect the one in~\cite{Espriu:2012ih}, since for a 
large portion of the $a_{4,5}$ parameter space many resonances have negative widths breaking causality.

It is usually expected that a new strongly interacting sector would lead to dynamical 
resonances in different channels 
but what turned out to be a bit of a surprise in our previous work~\cite{Espriu:2012ih} and in the present for $a<1$ 
is that these resonances are typically  narrow and very hard to detect. This appears
to be directly related to the unitarization of the $W_LW_L$ scattering in the presence of light Higgs.
Searching for these dynamical resonances at LHC  will be however very important 
because if none of them reveals itself below $\sim 3$ TeV
virtually all $a_{4,5}$ parameter space of the anomalous couplings could be excluded. This can be an indirect way of
assessing these quartic electroweak boson couplings. Actually, no direct information  on $a_4$ and $a_5$ exists 
at present from direct measurements of the quartic electroweak boson couplings.   

Unfortunately the actual signal strength of the new resonances predicted is such that they are not currently 
being probed in 
LHC Higgs search data and consequently no relevant bounds on $a_4$ and $a_5$ can 
be derived at present from the existing data ---a situation that may change soon. 
The previous considerations emphasize the importance of indirect measures of the couplings $a_4$ and $a_5$ 
by searching for the additional resonances coming out  from our study of $WLWL$ scattering.
Measuring these anomalous couplings will be one of the main tasks of the LHC run starting in 2015.

\section{Isospin and partial wave amplitudes}
Here we introduce the basic definition of our observables. We shall consistently assume our treatment that custodial symmetry is
exactly preserved. This implies taking $g^\prime=0$ and ignoring all the $\mathcal{O}_i$ 
operators that can contribute to $WW$ scattering but
$\mathcal{O}_4$ and $\mathcal{O}_5$. This approximation 
also allows for a neat usage of the isospin formalism and for the convergence
of the partial wave amplitudes. We also disregard operators that contain matter fields as 
they are totally irrelevant for the present discussion.

As emphasized in \cite{Espriu:2012ih} when dealing with longitudinally polarized amplitudes,
as opposed to using the ET approximation, caution must be exercised to account 
for an ambiguity introduced by the longitudinal polarization vectors that do not transform under
Lorentz transformations as 4-vectors.   
Expressions involving the polarization vector $\epsilon_{L}^\mu$ can not be cast in terms of 
the Mandlestam variables $s$, $t$, and $u$ 
until an explicit reference frame has been chosen, as they can not themselves be written 
solely in terms of covariant quantities. Obviously amplitudes still satisfy crossing 
symmetries when they remain expressed in terms of the external 4-momenta. A short discussion
on this point is placed in appendix~\ref{sec:appendix_crossing}.
 
A general  amplitude, $A(W^{a}(p^{a})+W^{b}(p^{b})\to W^{c}(p^{c}) + W^{d}(p^{d}))$,  can be written 
using isospin and Bose symmetries as 
\bea
A^{abcd}(p^{a},p^{b},p^{c},p^{d})&=&  \delta^{ab} \delta^{cd} A(p^{a},p^{b},p^{c},p^{d}) + 
\delta^{ac} \delta^{bd} A(p^{a},-p^{c},-p^{b},p^{d})\label{eq:isospin_general}\\
&+& 
\delta^{ad} \delta^{bc} A(p^{a},-p^{d},p^{c},-p^{b}),\no
\eea
with
\bea
\label{eq:amplitudes_general}
A^{+-00} & = & A(p^{a},p^{b},p^{c},p^{d}) \\ \no 
A^{+-+-} & = & A(p^{a},p^{b},p^{c},p^{d}) + A(p^{a},-p^{c},-p^{b},p^{d}) \\ 
A^{++++} & = & A(p^{a},-p^{c},-p^{b},p^{d}) + A(p^{a},-p^{d},p^{c},-p^{b}).\no 
\eea
The fixed-isospin amplitudes are given by
\bea
\label{eq:fixed_isospin}
T_{0}(s,t,u) & = & \langle 00\vert S\vert 00 \rangle  =   3 A^{+-00} +   A^{++++} \\ \no
T_{1}(s,t,u) & = & \langle 10\vert S\vert 10 \rangle  =   2 A^{+-+-} - 2 A^{+-00} - A^{++++} \\ \no
T_{2}(s,t,u) & = & \langle 20\vert S\vert 20 \rangle  =   A^{++++} \, . \nn 
\eea
We shall also need the amplitude for the process $W^+W^-\to hh$. Taking into account that the final state
is an isospin singlet and defining 
\be 
A^{+-}= A(W^{+}(p^{+})+W^{-}(p^{-})\to h(p^{c}) + h(p^{d})) \, ,
\ee
the projection of this amplitude to the $I=0$ channel gives 
\be
\label{eq:fixed_isospin_hh}
T_{H, 0}(s,t,u)= \sqrt{3} A^{+-}.
\ee
The partial wave amplitudes for fixed isospin $I$ 
and total angular momentum $J$ are 
\be\label{eq:legendre}
t_{IJ}(s) = \frac{1}{64\pi} \int_{-1}^{1} d(\cos\theta) P_{J}(\cos\theta) T_{I}(s,t,u) \, ,
\ee
where the $P_{J}(x)$ are the Legendre polynomials and $t = (1 - \cos\theta) (4 M^2-s)/2$,
 $u = (1 + \cos\theta) (4 M^2-s)/2$ with $M$ being the $W$ mass.  We will concern ourselves with only the lowest 
non-zero partial wave amplitude in each isospin channel: $t_{00}(s)$, $t_{11}(s)$, and $t_{20}(s)$, namely the 
scalar/isoscalar, vector/isovector, and isotensor amplitudes respectively. Unitarity directly implies that $|t_{IJ}(s)|<1$. 
For further implications of unitarity on  $t_{IJ}(s)$  the interested reader may see
ref.~\cite{unitarity}.  

In this work, the partial wave amplitude $t_{IJ}(s)$ are studied up to ${\cal O}(p^4)$, namely
\bea\label{eq:tloop}
t_{IJ}(s) = t_{IJ}^{(0)}(s) + t_{IJ}^{(2)}(s)\,.
\eea
Here $t_{IJ}^{(0),(2)}(s)$ are  tree-level and  ${\cal O}(p^4)$ contributions, respectively. 
$t_{IJ}^{(0)}(s)$  can be constructed from eq.~(\ref{eq:legendre}) 
by using crossing and isospin relation for the tree level contributions of $A^{+-00}$ (Figure~\ref{tree-diagrams}). 
The analytic results of $A^{+-00}$ at tree-level 
are in appendix~\ref{sec:appendix_amplitudes}.  $t_{IJ}^{(0)}(s)$ contains the anomalous coupling $a$ but $b$ does not enter
at tree-level.  $t_{IJ}^{(4)}(s)$  includes tree-level contributions from $a_i$  counter-terms 
(see appendix~\ref{sec:appendix_amplitudes} for analytic result) and 
the one-loop corrections to the diagrams  in Figure~\ref{tree-diagrams}. 
At one-loop level, the $b$ parameters enters $t_{IJ}^{(4)}(s)$ by the one-loop expression of $A^{+-00}$ calculated in ref.~\cite{EMY}.
\begin{figure}[h!]
\begin{center}
\includegraphics[clip,width=0.25\textwidth]{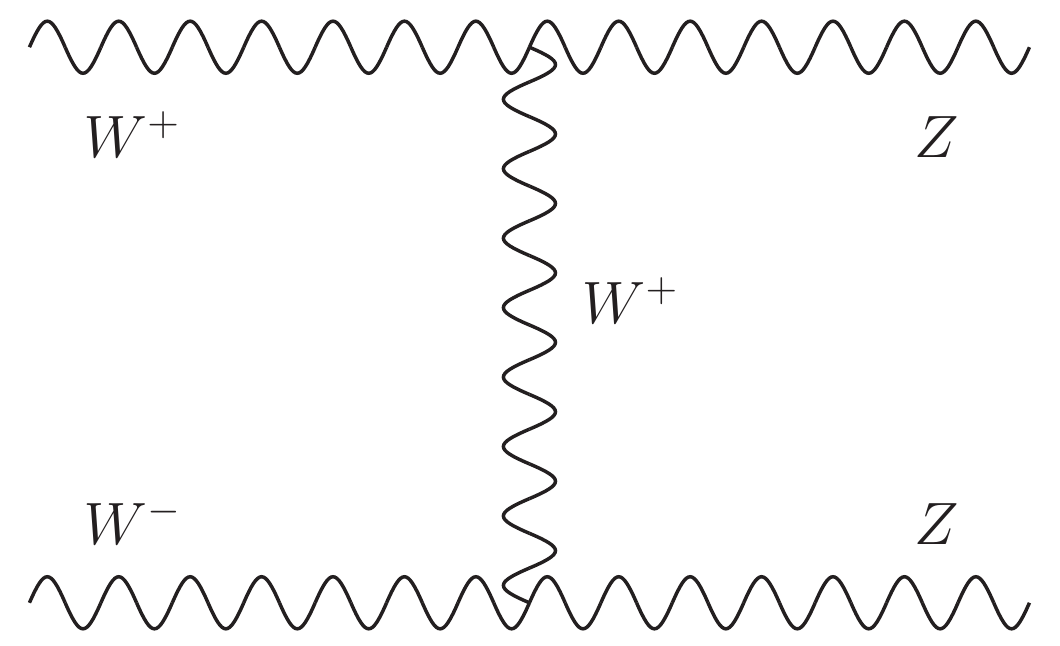} \hspace{0.70cm}
\includegraphics[clip,width=0.25\textwidth]{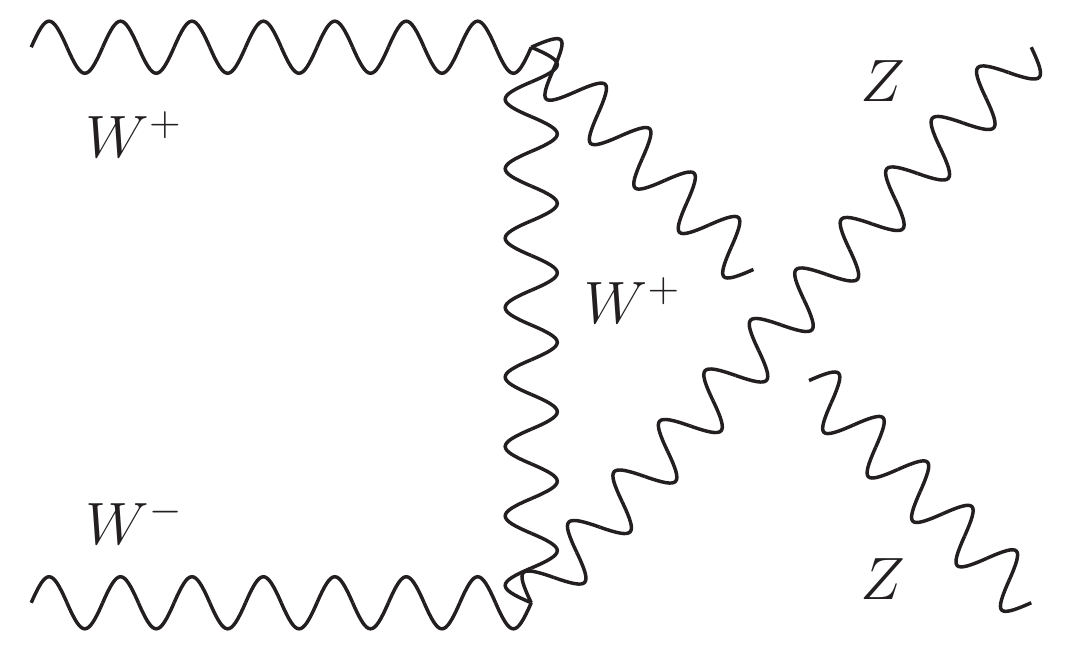} \\ \vspace{0.15cm}
\includegraphics[clip,width=0.25\textwidth]{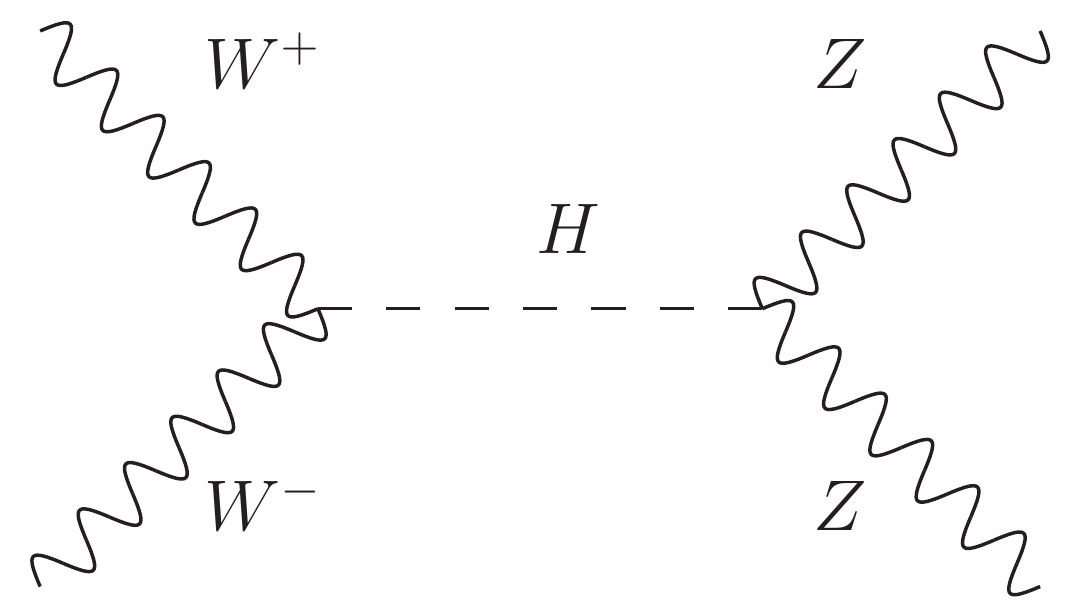} \hspace{0.70cm}
\includegraphics[clip,width=0.25\textwidth]{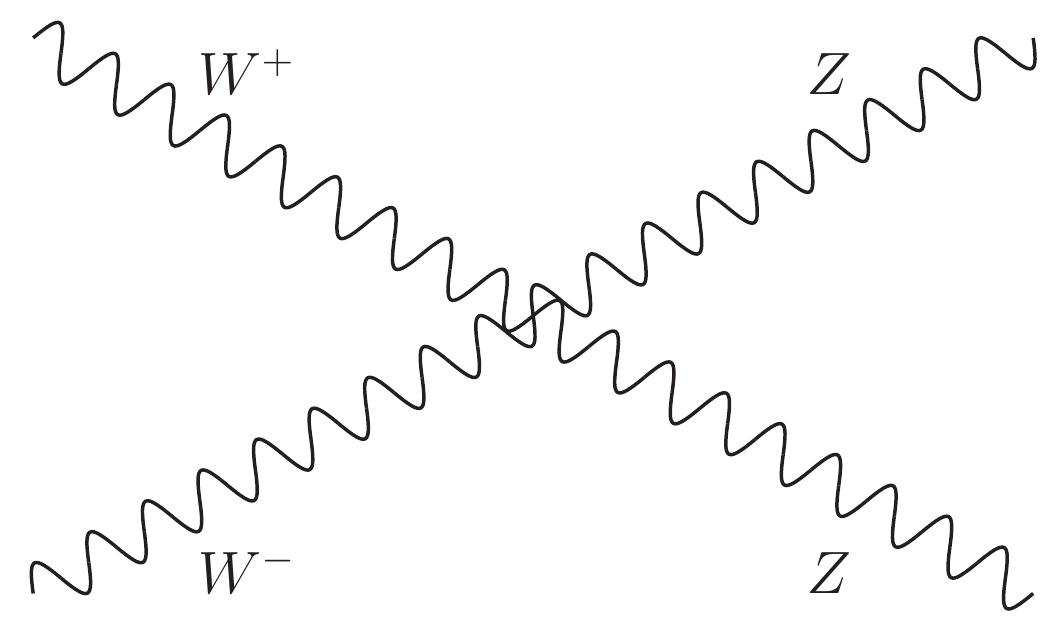}
\caption{Diagrams contributing to $A(s,t,u)$ at tree level.\label{tree-diagrams}}
\end{center}
\end{figure}

\section{Scrutiny of the tree-level amplitudes  $t_{00}^{(0)}$,  $t_{20}^{(0)}$ and  $t_{11}^{(0)}$ }
For values of $a$ different from $1$,  
the $W_LW_L$ scattering amplitudes exhibit rather different behaviour with respect to the MSM case $a=1$.
The most important difference is that the $|t_{IJ}|<1$ unitarity bound is violated at tree-level pretty quickly.  
We shall see later how to restore unitarity with the help of higher loops and counter-terms but in this section 
we concentrate on the peculiarities of the tree level amplitudes $t_{00}^{(0)}$,  $t_{20}^{(0)}$ and  $t_{11}^{(0)}$.
Here the partial wave amplitudes are studied in the complete theory, namely away from the ET approximation.
This is a key point since there are interesting kinematical features of  $t_{IJ}^{(0)}$ that are totally missed 
in the ET approximation, such as the presence of sub-threshold singularities and zeroes of  $t_{IJ}^{(0)}$ 
absent in ET approximation. Some of these features will be crucial in our analysis.

In order to study the behaviour of  $t_{IJ}^{(0)}$, 
we will establish three different regions according to the range of the values of $a=1$, $a>1$ and $a<1$.


\subsection{Case $a=1$} 
\begin{figure}[h!]
\begin{center}
\includegraphics[scale=0.85]{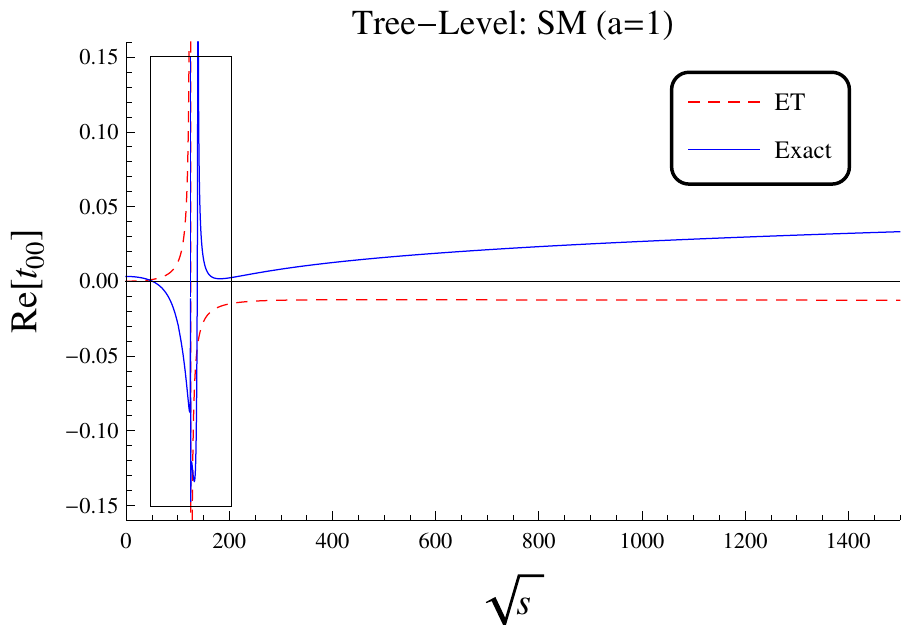}
\includegraphics[scale=0.85]{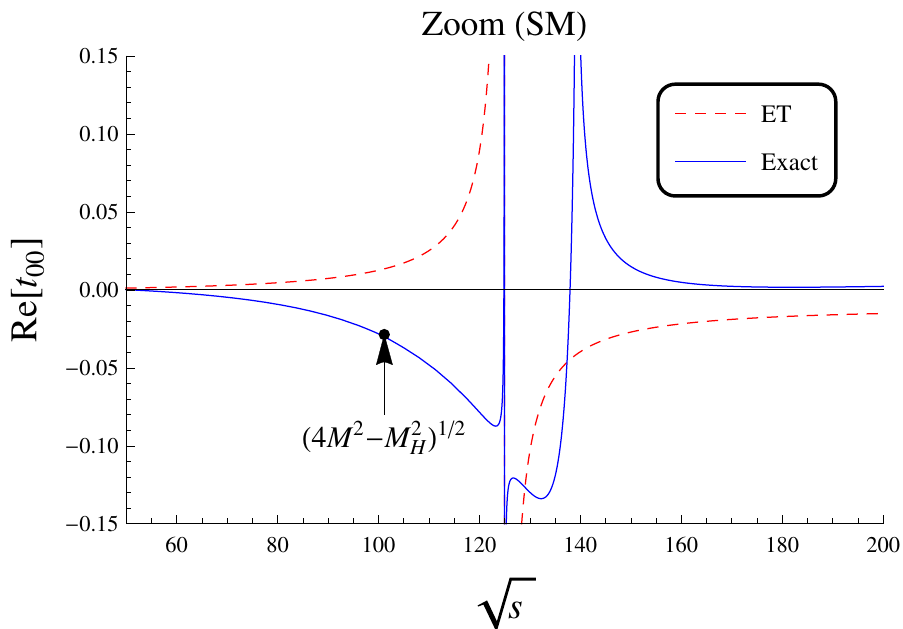}\\
\includegraphics[scale=0.85]{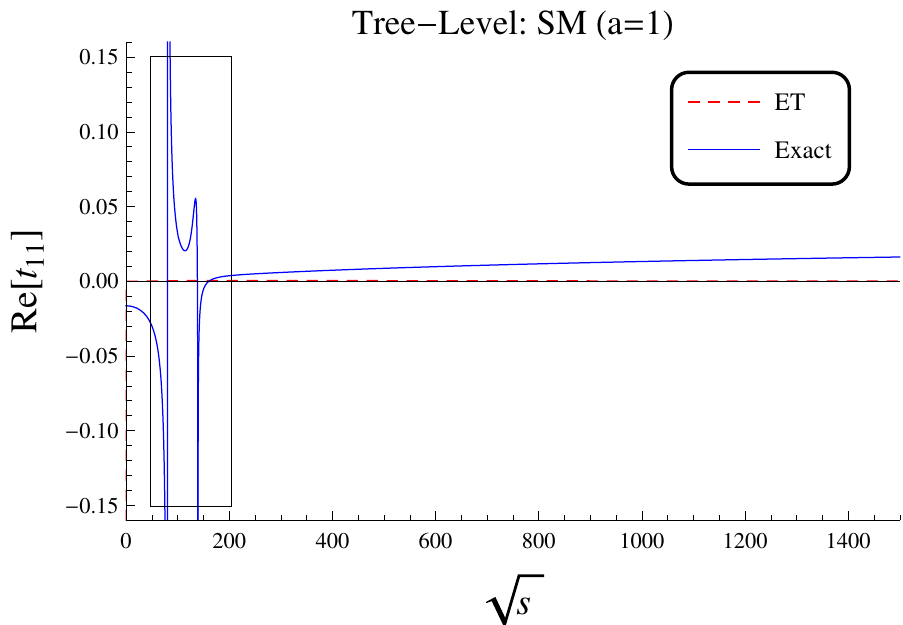}
\includegraphics[scale=0.85]{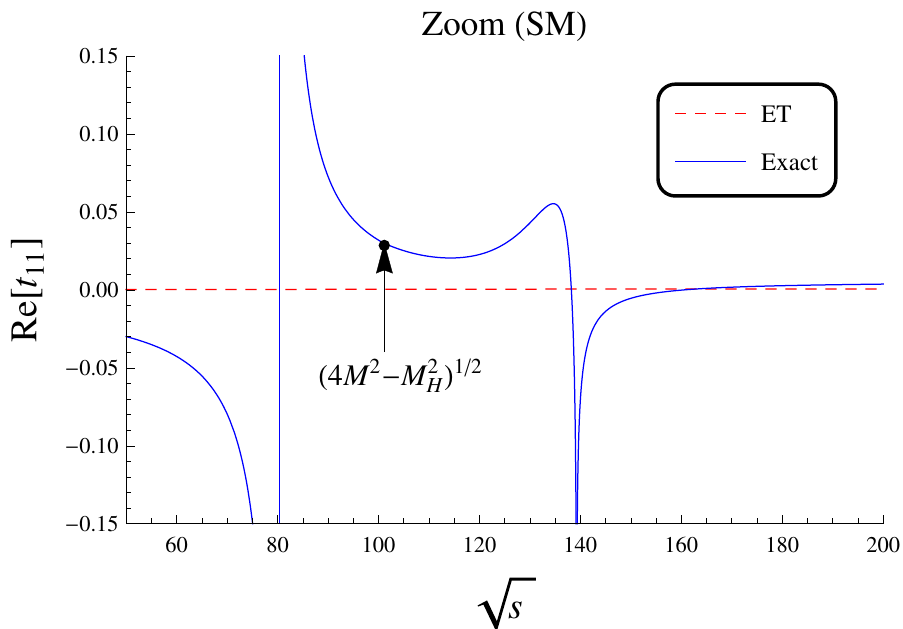}
\caption{Plot of $t_{00}^{(0)}$ (above) and $t_{11}$ (below) for $a=1$. In both cases a zoom on the lowest
values of $s$ to show the complete analytic structure is presented. The arrow indicates the position
of one of the sub-threshold singularities that is invisible at the scale of the
plot.\label{t00-SM}}
\end{center}
\end{figure}

In Figure~\ref{t00-SM}
we plot the tree-level  isoscalar partial wave amplitude $t_{00}^{(0)}(s)$ for $W_LW_L\to Z_LZ_L$ as a function of $s$. 
The external $W$ legs are taken on-shell ($p^2=M^2=M_W^2=M_Z^2$). 
As we see from Figure~\ref{t00-SM} the partial wave amplitude has a rather rich analytic structure. 
It has one pole at $s=M_H^2$ but also a second singularity can be seen 
at the value $s= 3M^2$. A closer examination reveals also a third singularity at $s= 4M^2-M_H^2$, 
invisible in the  Figure~\ref{t00-SM} as it happens
to be multiplied by a very small number. These singularities correspond to poles 
of the $t$ and $u$ channel diagrams in Figure~\ref{tree-diagrams} that after 
the angular integration of eq.~(\ref{eq:legendre}) to obtain the partial wave
amplitudes behave as logarithmic  divergences. The $t$ and $u$ channels are absent in the ET approximation. 
Note that both singularities are below the physical threshold at $s=4M^2$. Beyond the $s=3M^2$ singularity 
the amplitude for $a=1$ is always positive as can be seen in Figure \ref{t00-SM}. 
 
In Figure~\ref{t00-SM} we also plot the tree-level partial wave amplitude $t_{11}^{(0)}(s)$. 
Here, a pole at $s=M^2$ is visible, as expected, along with the two kinematical sub-threshold singularities
already mentioned. In Figure~\ref{t00-SM} the $t_{00}^{(0)}$ and $t_{11}^{(0)}$ amplitudes 
are also compared with the respective amplitudes obtained in ET approximation (computed assuming $M=0$
as is customary).
As can be seen the ET is grossly inadequate at low energies. In particular it fails in 
reproducing the rich analytic structure of the amplitudes. 
The non-analyticity at $s= 3M^2$  and $s= 4M^2-M_H^2$ due to sub-threshold singularities
is actually also present in the $t_{20}^{(0)}$ partial 
wave amplitude (not depicted), corresponding like in the other two cases to a (zero width) logarithmic 
pole. In $t_{20}^{(0)}$ there are no other singularities as
no $I=2$ particle is exchanged in the $s$-channel.
These sub-threshold singularities are genuine effects in the $W_LW_L\to ZZ$ amplitudes 
and are independent from the value of $a$.  
These features  are conspicuously absent in the analogous amplitude computed in the ET.

$W_LW_L\to Z_LZ_L$ scattering can be accessible at LHC by the studying the process $pp\to WW jj$.
Then, these sub-threshold singularities should be hardly visible mostly due to the off-shellness of
the $W_LW_L\to Z_LZ_L$ amplitude on $pp\to WW jj$.   The experimental process spreads the logarithmic poles over a range of 
invariant masses. For instance, the singularity at $s=M^2$ appears actually
at $s=\sum q_i^2 -M^2$ if $W$ legs are off-shell. In addition cuts in $p_T$ should render the partial wave amplitude 
actually non-singular~\footnote{We thank D. D'Enterria and X. Planells for discussions on these points.}.


\subsection{Case $a>1$}
\begin{figure}[h!]
\begin{center}
\includegraphics[scale=0.85]{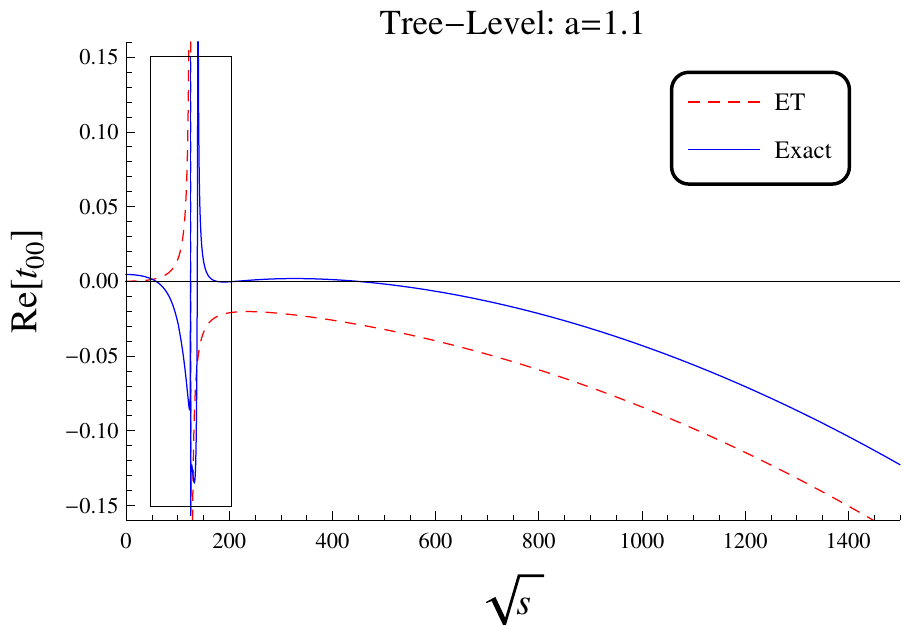}
\includegraphics[scale=0.85]{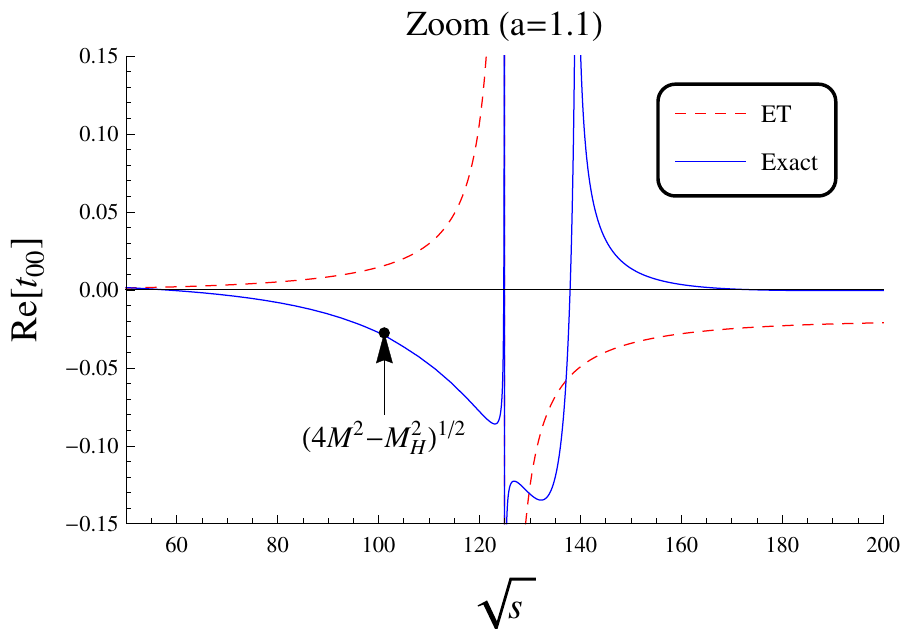}\\
\includegraphics[scale=0.85]{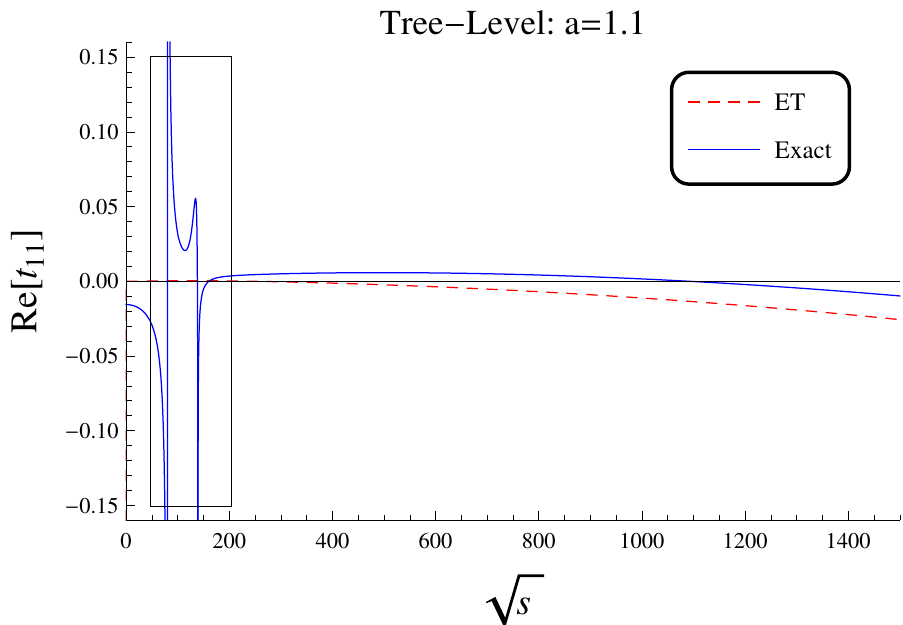}
\includegraphics[scale=0.85]{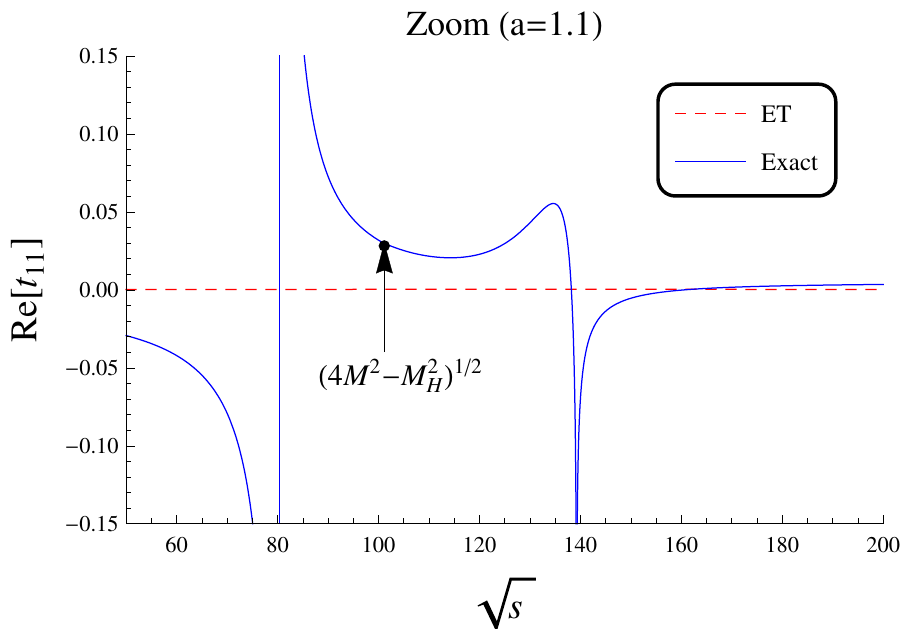}
\caption{Plot of $t_{00}^{(0)}$ and $t_{11}^{(0)}$ for $a=1.1$ and a zoom on 
the low $s$ region where the amplitude is very small. Several additional zeroes appear above threshold
and is not unitary. The ET result is shown by (red) a dotted line.\label{t00-a11}}
\end{center}
\end{figure}

\begin{figure}[h!]
\begin{center}
\includegraphics[scale=0.85]{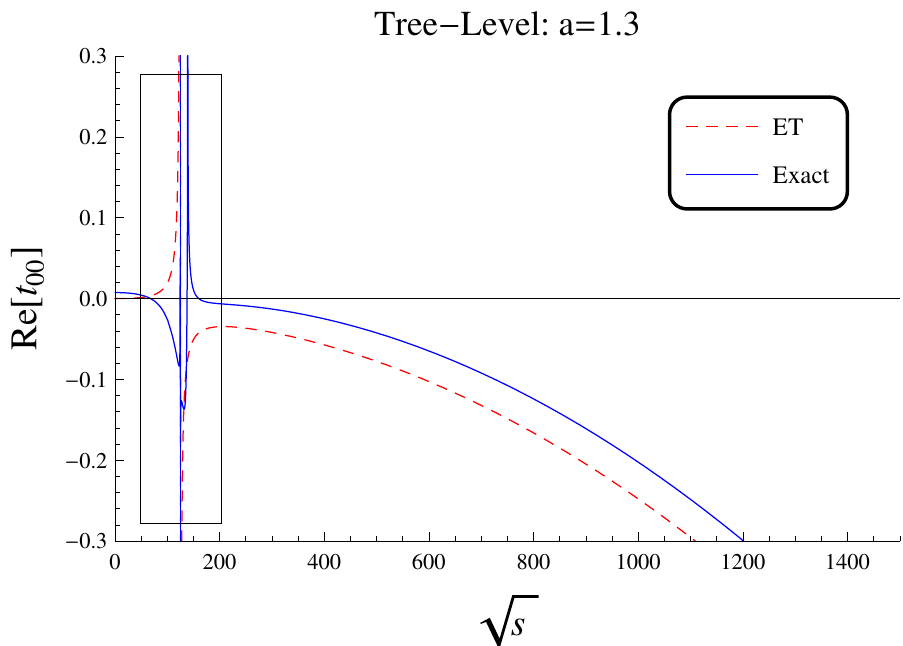}
\includegraphics[scale=0.85]{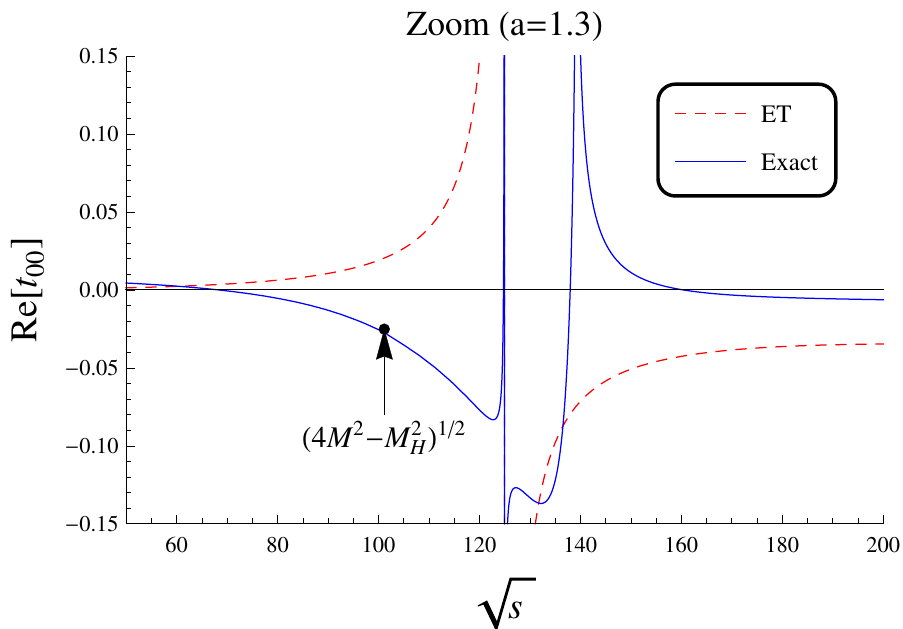}\\
\includegraphics[scale=0.85]{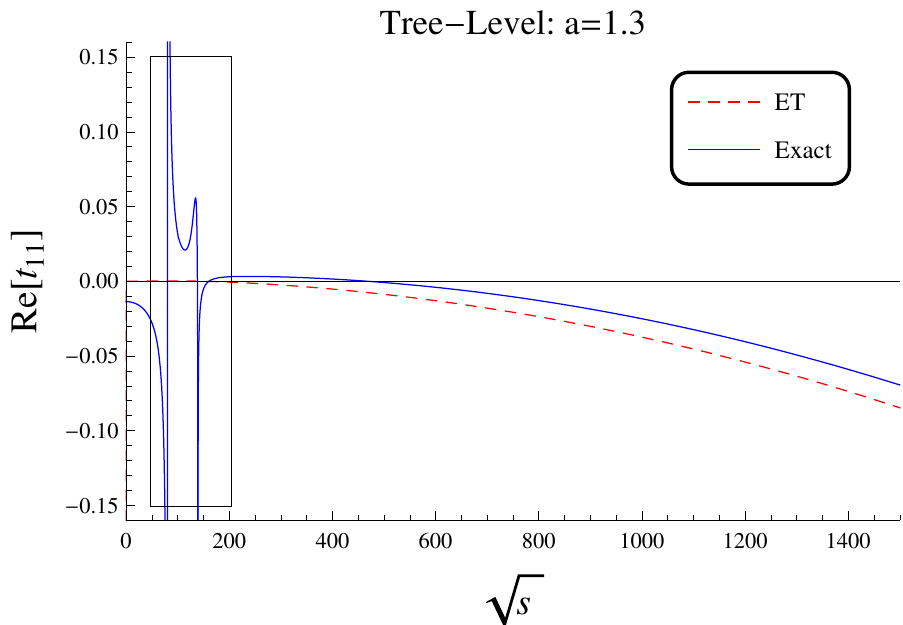}
\includegraphics[scale=0.85]{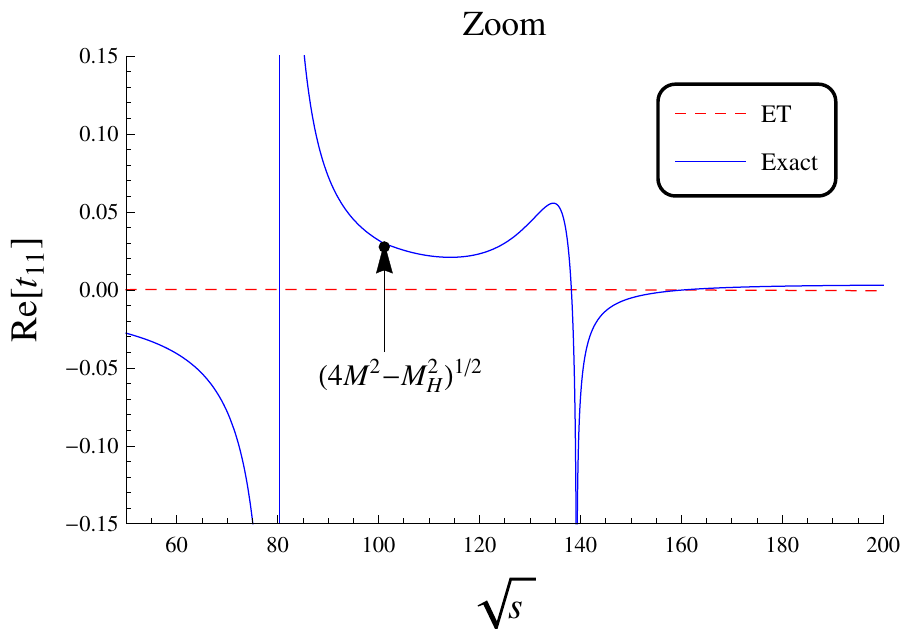}
\caption{Plot of $t_{00}^{(0)}$ and $t_{11}^{(0)}$ for $a=1.3$. Amplitudes are not unitarity. The three singularities commented on the text 
are shown in each channel.  Amplitudes are not vanishing 
for valeus of $s$  up to the range of validity of the effective Lagrangian. The ET results are
indicated by a dotted line.\label{t00-a13}}
\end{center}
\end{figure}

The three sub-threshold singularities appearing at $a=1$  are also present in this case.
However, for $a>1$ the partial wave amplitudes also show a new features. First of all, as shown
in Figure~\ref{t00-a11} for $a=1.1$ and amplitudes $t_{00}^{(0)}(s)$ and $t_{11}^{(0)}(s)$, 
the tree-level partial wave amplitude for $t_{IJ}^{(0)}(s)$  show 
clear non-unitary behaviours as it goes to $-\infty$ 
as $s$ increases. In addition, for $a>1$ 
the tree-level partial wave amplitudes for $t_{IJ}^{(0)}(s)$  have zeroes for values of $s$ above threshold and well below 
well below the cut-off scale ($3$ TeV) of our effective Lagrangian.
Setting for example the value $a=1.1$ compatible with the experimental constraint in eq.~(\ref{eq:bound_a_a4_a5}) 
the  $t_{00}^{(0)}(s)$ amplitudes vanishes at two values of $\sqrt{s}$ around $216$ and $445$ GeV 
(see Figure~\ref{t00-a11} for $a=1.1$),
the  $t_{11}^{(0)}(s)$ at a value around $1$ TeV as well as  $t_{20}^{(0)}(s)$ at  about $800$ GeV (not shown). 
For $a>1.125$,  the tree-level amplitude $t_{00}^{(0)}$ has no zeroes (Figure~\ref{t00-a13} for $a=1.3$), whereas
the $t_{11}^{(0)}(s)$ and $t_{20}^{(0)}(s)$amplitudes for values of $a$ compatible 
with bounds in eq.~(\ref{eq:bound_a_a4_a5}) still vanish at specific values of $\sqrt s$. 
For example for $a=1.3$, 
the zeroes of $t_{11}^{(0)}(s)$ and $t_{20}^{(0)}(s)$  
are at $\sqrt{s}$ around $450$ GeV. The presence of zeroes for the tree-level amplitudes 
at low values of $\sqrt{s}$ is interesting point as it means that around these zeroes the $W_LW_L\to Z_LZ_L$ amplitudes 
are strongly suppressed. 
It may be relevant to note that the $t_{00}^{(0)}$ and $t_{11}^{(0)}$  amplitudes are very small over a fairly extended
range of values of $s$ for a range of values of $a>1$ (particularly so in the isovector channel). These facts
could perhaps be used to set rather direct bounds on this particular coupling. This issue deserves further
phenomenological study.  

\subsection{Case $a<1$}
\begin{figure}[h!]
\begin{center}
\includegraphics[scale=0.85]{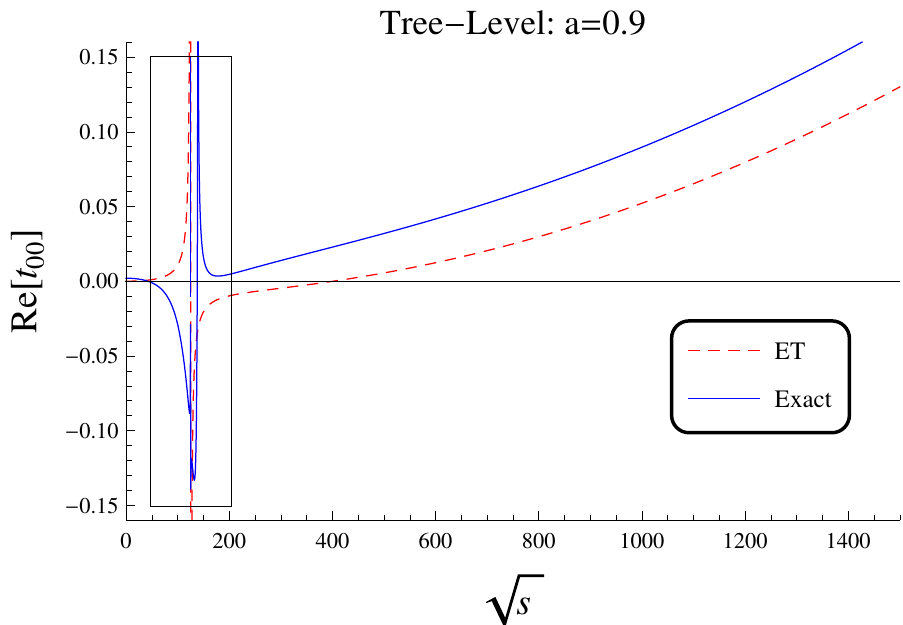}
\includegraphics[scale=0.85]{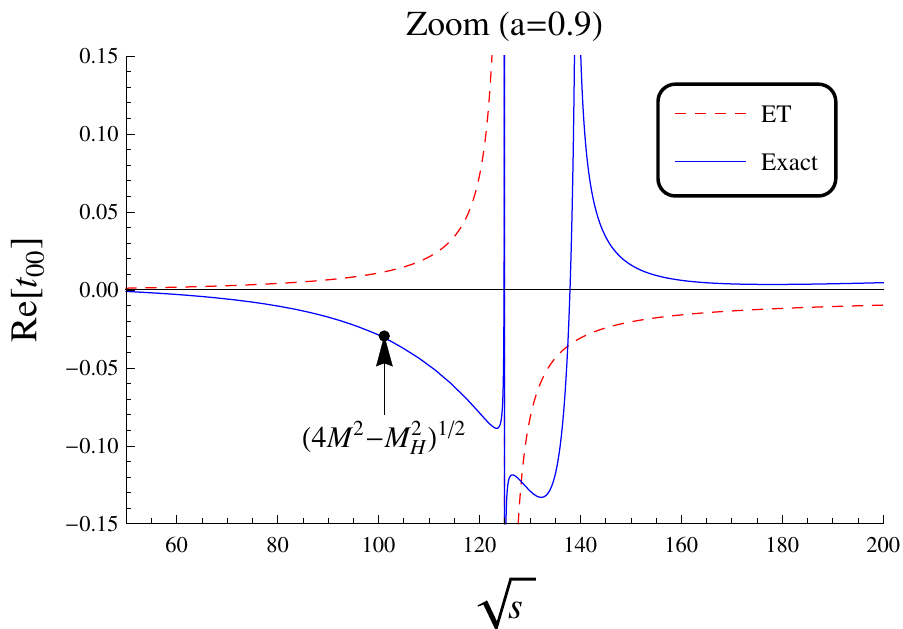}\\
\includegraphics[scale=0.85]{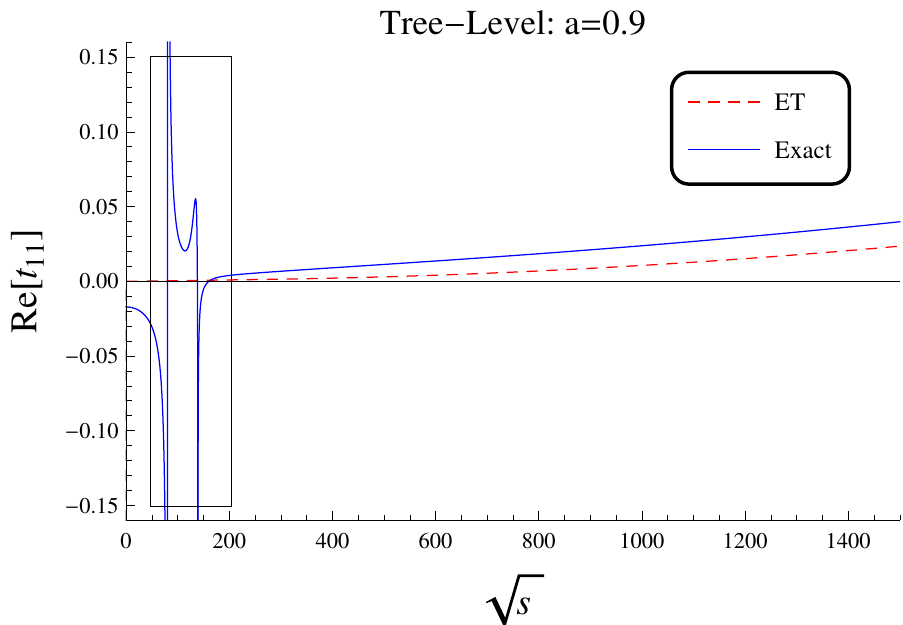}
\includegraphics[scale=0.85]{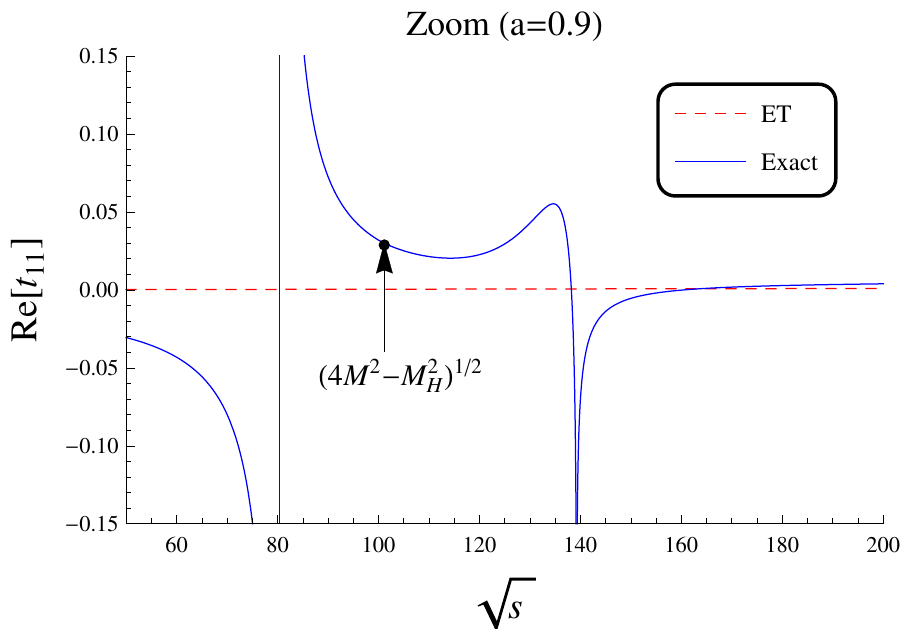}
\caption{Plots of $t_{00}^{(0)}$ and $t_{11}^{(0)}$ for $a=0.9$ and a zoom of the region
at low $s$ where the amplitudes are very small. No additional zeroes appear
and the amplitudes also show a non-unitary behaviour at large $s$. The nearly
invisible logarithmic singularity at $s=4M^2-M_H^2$ is indicated. The results in the ET 
approximation are also indicated by a dotted line.\label{t00-a09}}
\end{center}
\end{figure}

For $a<1$, the $t_{IJ}^{(0)}$ amplitude still present the two sub-threshold 
singularities at $s= 3M^2$  and $s= 4M^2-M_H^2$. Beyond them however,   
no additional zeroes appear, amplitudes are positive and go to $\infty$ as $s$ increases.  
This clearly reflects the non-unitary character of  $t_{IJ}^{(0)}$ amplitudes for $a\ne1$. 
In Figure~\ref{t00-a09}, we show as an example the 
$t_{11}^{(0)}(s)$ and $t_{20}^{(0)}(s)$ amplitudes in the case $a=0.9$.
The equivalent amplitudes computed by making use of the ET are also shown in Figure~\ref{t00-a09}. 
Both in this $a<1$ case and in the $a>1$ one we see that the ET works reasonably well
for large values of $s$, but again fails at low and moderate values.

\section{Unitarity corrections}
In the case of Higgs anomalous couplings to gauge sector ($a\neq 1$ and $b\ne1$) 
the tree-level amplitudes $t_{IJ}^{(0)}$ are not-unitarity and we are forced to include 
additional operators in the theory, such as the 
$a_i$ counter-terms  in eq.~(\ref{eq:1}). At one-loop level, the $a_{i}$ will cancel 
the divergences of the Lagrangian in eq.~(\ref{eq:1}) and finite 
couplings renormalized at some UV scale will remain~\cite{EMY,dobado}, namely 
\bea
a_4\vert_{\rm finite} &\simeq& \frac{1}{(4\pi)^2}\frac{-1}{12} (1-a^2)^2  \log\frac{v^2}{f^2}
\\
a_5\vert_{\rm finite} &\simeq& \frac{1}{(4\pi)^2}\frac{-1}{24} \left[(1-a^2)^2+
\frac32((1-a^2)-(1-b))^2\right] \log\frac{v^2}{f^2},
\eea
where $f$ is the scale of the new interactions, and possibly other finite pieces. 

Up to now, the calculation of the one-loop $t_{IJ}^{(2)}(s)$ contribution in eq.~(\ref{eq:tloop}) 
is not available for $a$ and $b$ arbitrary and longitudinally polarized $W$ and $Z$. This would require the
evaluation of over one thousand diagrams. A numerical calculation is only available in~\cite{denner} 
for the case $a=b=1$ but it is not very useful for our purposes.  

For this reason, to estimate the $t_{IJ}^{(2)}(s)$ contribution in eq.~(\ref{eq:tloop}) we proceed in the following way.
The analytic contribution from $a_{4,5}$ terms are calculated exactly with  longitudinally polarized $W$ and $Z$ 
(appendix~\ref{sec:appendix_amplitudes}) like the tree-level contribution $t_{IJ}^{(2)}(s)$ . 
The real part of $t_{IJ}^{(2)}(s)$ will however be determined using the ET~\cite{ET,esma}; i.e. 
we replace this loop amplitude by the corresponding process $w^+ w^- \to zz$.  
For this part of the calculation we take $q^2=0$ for external legs and set $M=0$ but the Higgs mass 
is kept. 
The relevant diagrams of $A(ww \to zz)$ entering  $t_{IJ}^{(4)}(s)$ 
were calculated in~\cite{EMY} where explicit expressions for the different 
diagrams for arbitrary values of the 
couplings $a$ and $b$ can be found. This calculation has been checked and extended in~\cite{dobado}, albeit
setting $M_H=0$. 
As to the imaginary part of $t_{IJ}^{(2)}(s)$ we can take advantage of the optical theorem to circumvent 
the problem of using the ET approximation.
In the $I=1, J=1$ and $I=2,J=0$ cases we can use the relations
\be
\label{eq:part_unitarity}
{\rm Im \,} t_{IJ}^{(2)}(s) = \sigma(s) |t_{IJ}^{(0)}(s)|^{2} \, ,
\ee
While for the $I=0$ amplitude we also have  a contribution from a two-Higgs
intermediate state. Then
\be
\label{eq:full_unitarity_hh}
{\rm Im \,} t_{00}(s) = \sigma(s) |t_{00}(s)|^{2} + \sigma_{H}(s) |t_{H, 0}(s)|^{2} \, ,
\ee
with
\be
\sigma(s) = \sqrt{1 - \frac{4 M^{2}}{s}} \hspace{1cm},\hspace{1cm}  \sigma_{H}(s) = \sqrt{1 - \frac{4 M_{H}^{2}}{s}}.
\ee
We believe that for the purpose of identifying dynamical resonances, normally occurring at $s\gg M_H^2$ the approximation
of relying on the ET for the real part of the loops is fine. Note that the dominant contribution 
to the real part for large $s$, of order $s^2$, is controlled 
by the contribution coming from couplings $a_{4,5}$. We have also actually
checked that, unless $a_4$ and $a_5$ are both very small, the contribution from the real part of the loop
amounts only to a small correction to $t_{IJ}^{(2)}$.

The final ingredient we need is a procedure to construct an unitary amplitude that perturbatively
coincides with the tree plus one-loop result but incorporates the principle of unitarity. 
To this purpose, we use the Inverse Amplitude Method (IAM)~\cite{iam} to the amplitude in eq.(\ref{eq:tloop}), namely 
\be
\label{eq:t_iam}
t_{IJ} \approx \frac{t_{IJ}^{(0)}}{1-t_{IJ}^{(2)}/t_{IJ}^{(0)}},
\ee
which is identical to the [1,1] Pad\'e approximant to $t_{IJ}$ derived from (\ref{eq:tloop}). 
The above expression obviously reproduces the first two orders of the perturbative expansion (eq.~\ref{eq:tloop})and, in addition, 
satisfies the necessary unitarity constraints, namely $|t_{IJ}|<1$ at high energies and 
\be
\label{eq:full_unitarity}
{\rm Im \,} t_{IJ}(s) = \sigma(s) |t_{IJ}(s)|^{2},
\ee
when the perturbative ingredients satisfy
\be
\label{eq:part_unitarity2}
{\rm Im \,} t_{IJ}^{(2)}(s) = \sigma(s) |t_{IJ}^{(0)}(s)|^{2} \, ,
\ee
as they must from the optical theorem.  We refer
to~\cite{Espriu:2012ih} and references therein for a more detailed discussion. We also recommend to read the
recent article \cite{ddll} for a rather complete review. In what follows we shall adhere to
the procedure outlined in \cite{Espriu:2012ih}.

There is no really unambiguous way of applying the IAM to the case where there are coupled channels
with different thresholds.
This will be relevant to us only in the $t_{00}$ case as there is an intermediate state consisting
in two Higgs particles. Here we shall adhere to the simplest choice that
consists in assuming  (\ref{eq:t_iam}) to remain valid also in this case. In addition, there
is decoupling of the two $I=0$ channels in the case $a^2=b$, as also discussed in \cite{ddll} in the context of the 
ET approximation. We have checked our results for different values of $b$, in particular we see that
setting $b=a^2$ does not give for the resonances that are eventually found results that are
noticeably different from those obtained for other values of $b$. Finally, we have also checked explicitly 
the unitarity of our results.

\section{Looking for resonances in $a_4$ and $a_5$ parameter space} 
Non-renormalizable models such as the effective theory described by the Lagrangian (\ref{eq:1})
typically produce scattering amplitudes that grow too fast with the scattering energy 
breaking the unitarity bounds~\cite{unitarity} at some point or other.

Chiral descriptions of QCD~\cite{GL} are archetypal examples of this behavior and unitarization 
techniques have to be used to recover unitarity. The  IAM~\cite{iam}, described in the previous section, 
is a convenient way of doing so. In QCD when
the physical value of the pion decay constant $f_\pi$ and the ${\cal O}(p^4)$ low energy 
terms $L_i$ (as defined e.g. in \cite{GL}, the counterpart of the $a_i$ in strong interactions) 
are inserted in the chiral Lagrangian and the IAM method is used,
the validity of the chiral expansion is considerably extended and one is able to reproduce the $\rho$ meson pole as 
well as many other properties of low energy QCD~\cite{iam}.
The limitations of the method derive to a large extent
from the accuracy in our knowledge of the different amplitudes entering the game. Different 
unitarization methods (such as e.g. N/D expansions or the Roy equations) always give very 
similar results as far as the first dynamical resonances is concerned.

Any strongly interacting theory should exhibit an infinite number of resonances. 
This is what hopefully one would get if all the terms
in the effective expansion were included, including all loop corrections and
counter-terms. Including contributions up to $\mathcal{O}(p^4)$,  our expression of $t_{IJ}(s)$ 
are to large extent  polynomials up to order $s^2$ (module logs). 
Therefore, we can find one or two resonances---the lowest lying ones
in each channel. However this is already providing us precious information on the dynamics
of the strongly interacting theory. In the present case, the
mere presence of higher dynamical resonances signals gives interesting 
information on the higher order coefficients of the effective 
Lagrangian (\ref{eq:1}) and therefore on $WW$ scattering.

If instead of a new strongly interacting sector the EWSBS is perturbative,
with point-like fields (a possibility could be an extended scalar sector or two Higgs-doublet models),
integrating them out would yield no-vanishing values for the coefficients $a_4$ and $a_5$~\cite{escia}. 
The unitarization method then reproduces approximately the masses of the particles that were originally 
integrated out which is still
valid information for physics beyond the SM.  
 
To find resonances, we perform a scan for the values
$|a_4|<0.01$ and $|a_5|<0.01$ and $a$ and $b$ fixed  looking for the
presence or otherwise of resonances in the different channels. We will consider
the different cases for $a\ne1$ since the case $a=1$ was discussed in detail in \cite{Espriu:2012ih}.

When looking for dynamical resonances we use two different methods. First we look for a zero of the real part
of the denominator in (\ref{eq:t_iam})
and use the optical theorem to determine the imaginary part ---i.e. the width--- at that location. A second method
consists in searching directly for a pole in the complex plane. 
In our case both methods give
very similar results, the reason being that the widths are typically quite small. It should be stated right
away that because of the way we compute the full amplitude, with separate derivations of the real and the
imaginary parts, the analytic continuation to the whole complex plane for $s$ is somewhat ambiguous. Had we found 
large imaginary parts some doubts could be cast on the results but fortunately this is not the case in virtually all 
of parameter space. Of course, a mathematical zero in the denominator (i.e. a genuine pole in the amplitude
$t_{IJ}$) is sometimes very difficult to get numerically, but proper resonances tend to reveal themselves in a 
rather clear way nevertheless. Some difficult cases present themselves for $a>1$ when the putative
resonance is close to one of the zeroes of the tree-level amplitude that appear in 
this case and we had to study these situations carefully. 

Physical resonances must have
a positive width and are only accepted as genuine resonances if $\Gamma < M/4$. 
Theories with resonances having a  negative width violate causality and the corresponding values of
the low energy constants in the effective theory  are to be rejected as leading to unphysical theories.
No meaningful microscopic theory could possibly lead to these values for the effective couplings.

\subsection{Case $a<1$} 
We start by considering this case where the unitarized amplitudes $t_{00}^{(0)}$, $t_{11}^{(0)}$ and  
$t_{20}^{(0)}$ share some properties with the ones from reference \cite{Espriu:2012ih} for 
$a=1$, namely the tree-level
amplitude has no zeros beyond the kinematical singularity existing
at $s= 3M^2$. In this case the sign of the tree-level amplitude as $s\to\infty$ is always positive in our conventions. 

\begin{figure}[tb]
\centering
\subfigure[(a)]{\includegraphics[clip,width=0.45\textwidth]{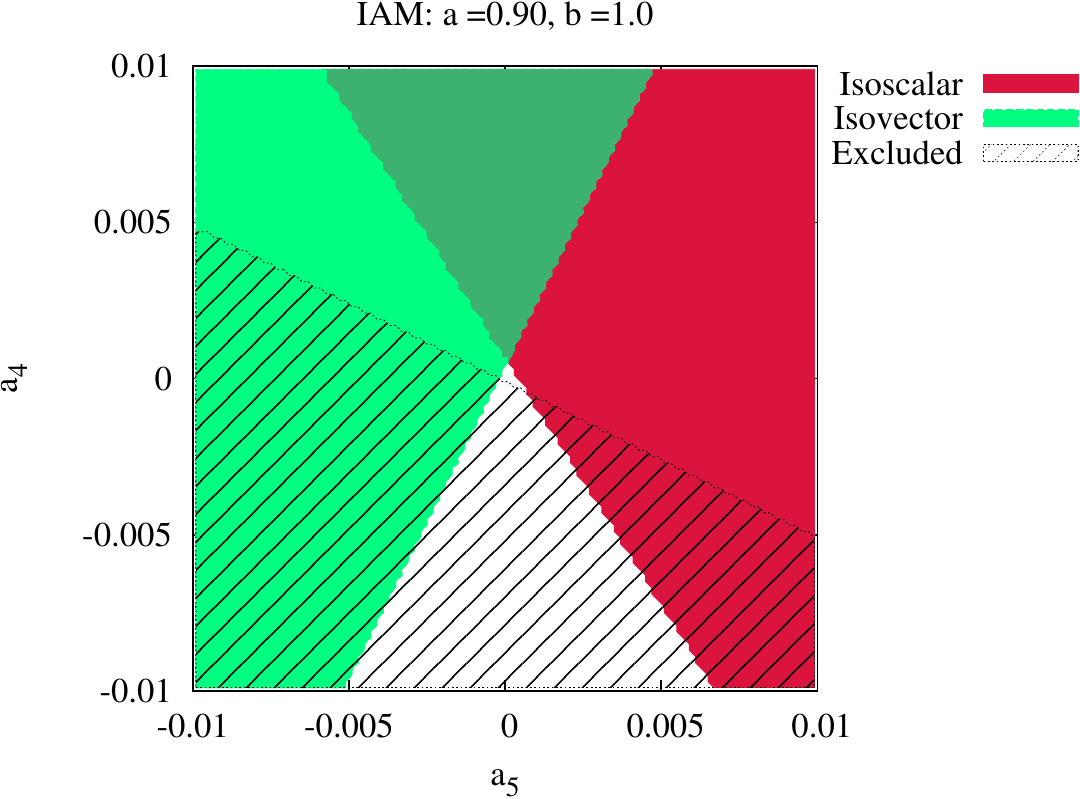}}
\subfigure[(b)]{\includegraphics[clip,width=0.45\textwidth]{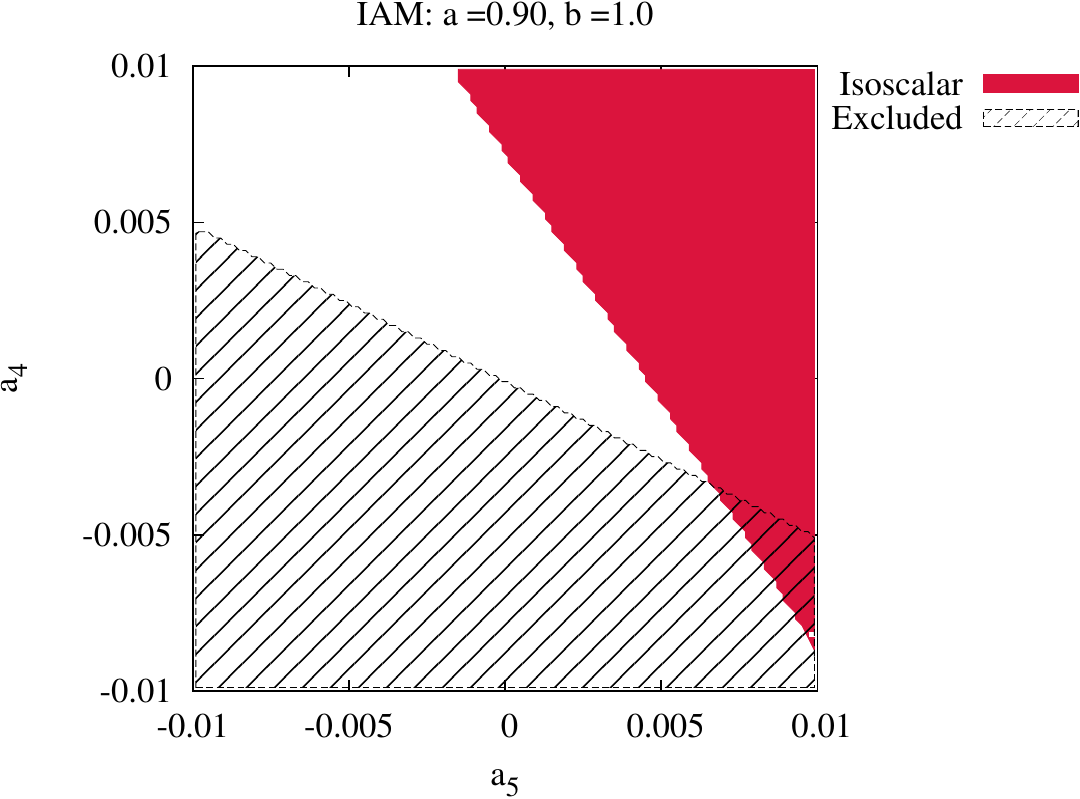}} \\
\caption{For $a=0.9$ and $b=1$: (a) Regions with isoscalar and isovector resonances
 (and the isotensor \textit{exclusion} region) up to a scale $4\pi v \approx 3$~TeV.  (b) Same as (a), 
but only showing isoscalar/isovector resonances in which $M_{S,V}<600$~GeV, for comparison with current 
Higgs search results.\label{explot09}}
\end{figure}

\begin{figure}[tb]
\centering
\subfigure[(a)]{\includegraphics[clip,width=0.45\textwidth]{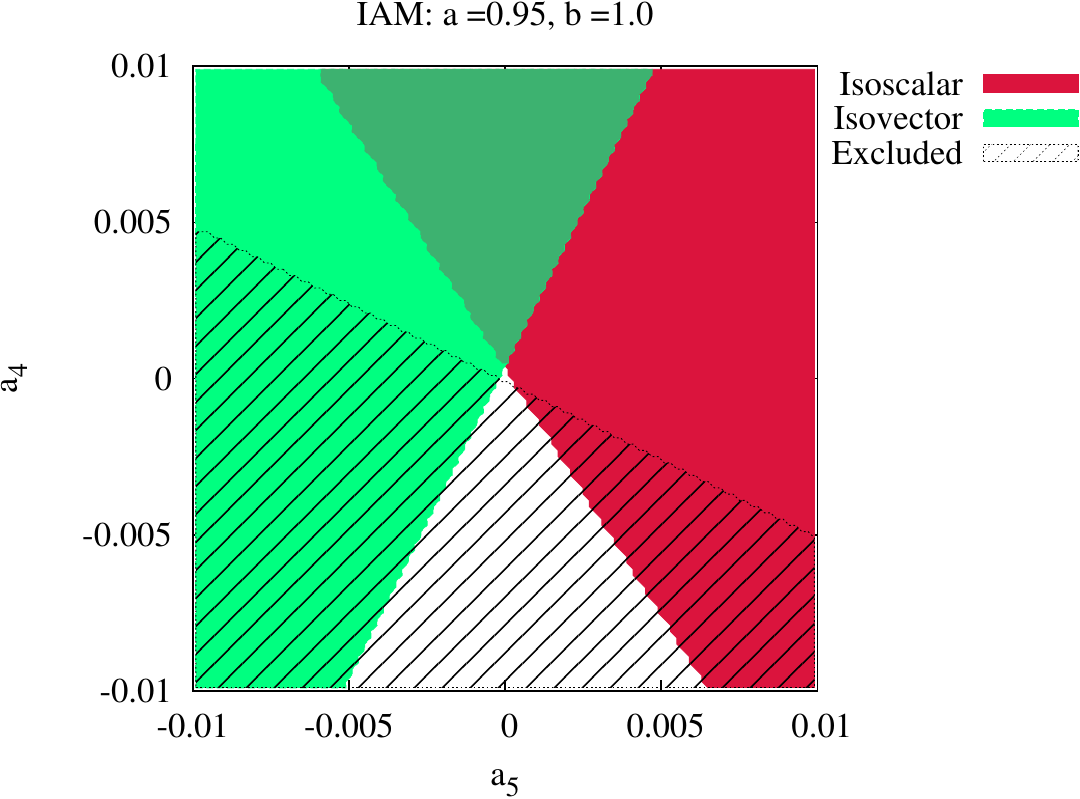}}
\subfigure[(b)]{\includegraphics[clip,width=0.45\textwidth]{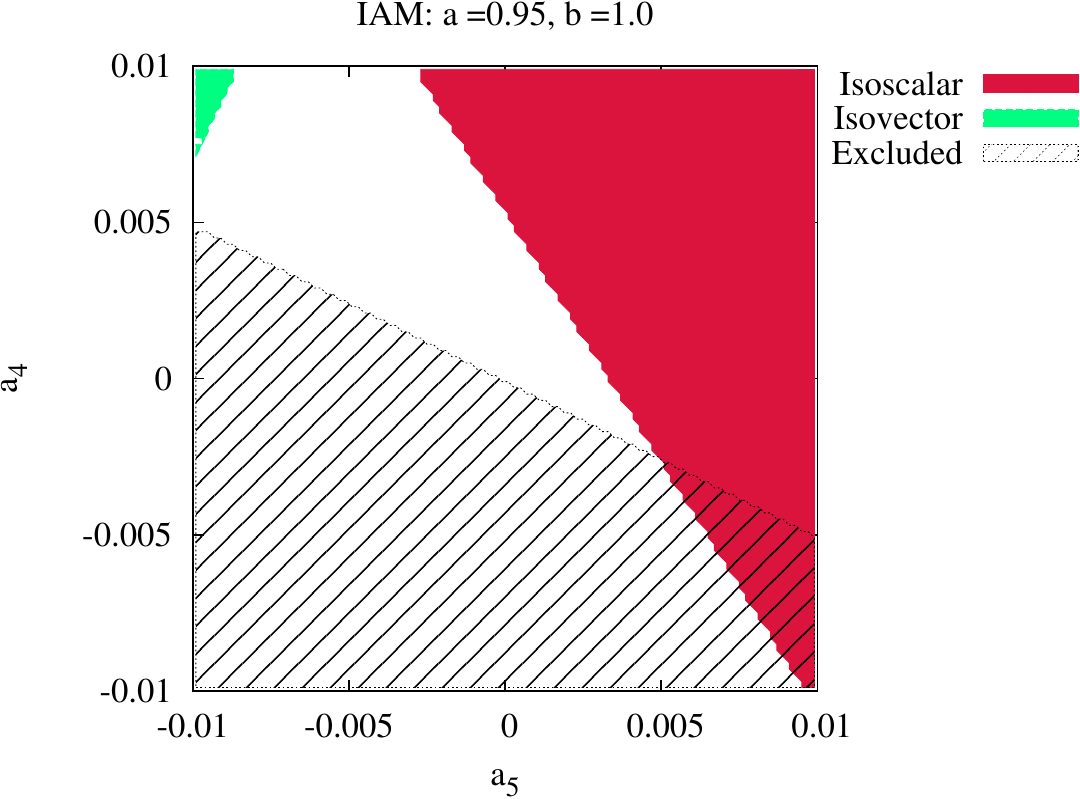}} \\
\caption{For $a=0.95$ and $b=1$:  (a) Regions with isoscalar and isovector resonances
 (and the isotensor \textit{exclusion} region) up to a scale $4\pi v \approx 3$~TeV.  (b) Same as (a), 
but only showing isoscalar/isovector resonances in which $M_{S,V}<600$~GeV, for comparison with current 
Higgs search results.\label{explot095}}
\end{figure}

\begin{figure}[tb]
\centering
\subfigure[(a)]{\includegraphics[clip,width=0.45\textwidth]{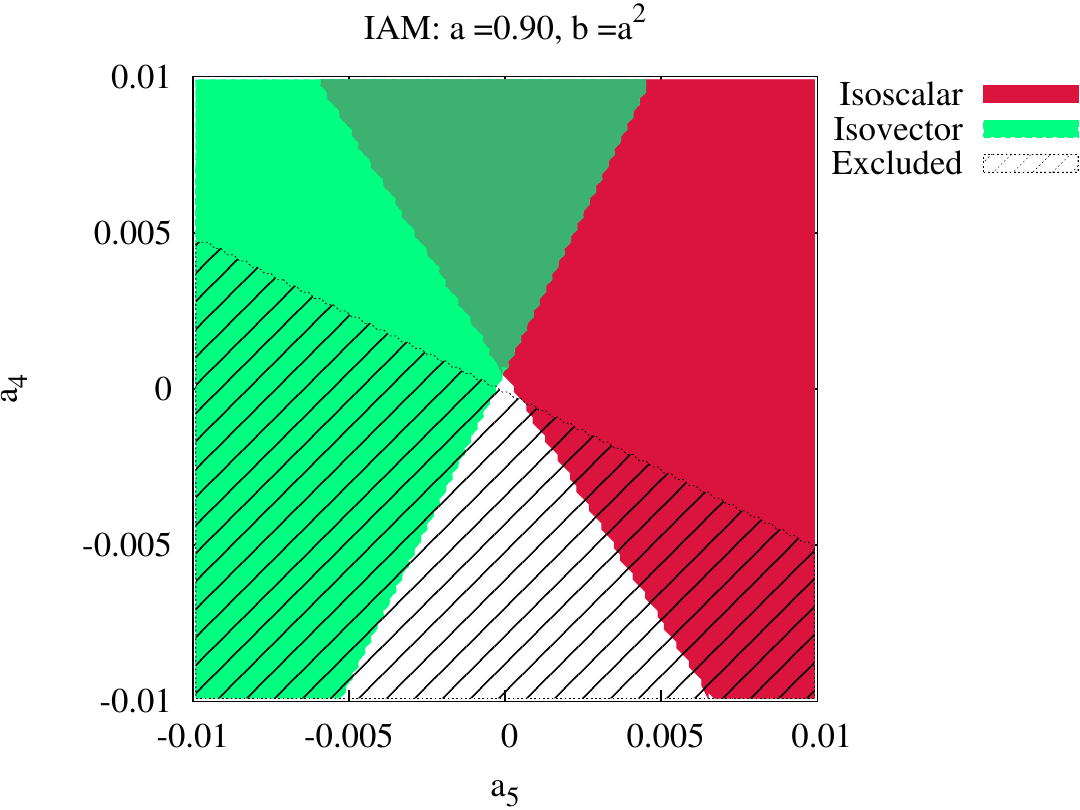}}
\subfigure[(b)]{\includegraphics[clip,width=0.45\textwidth]{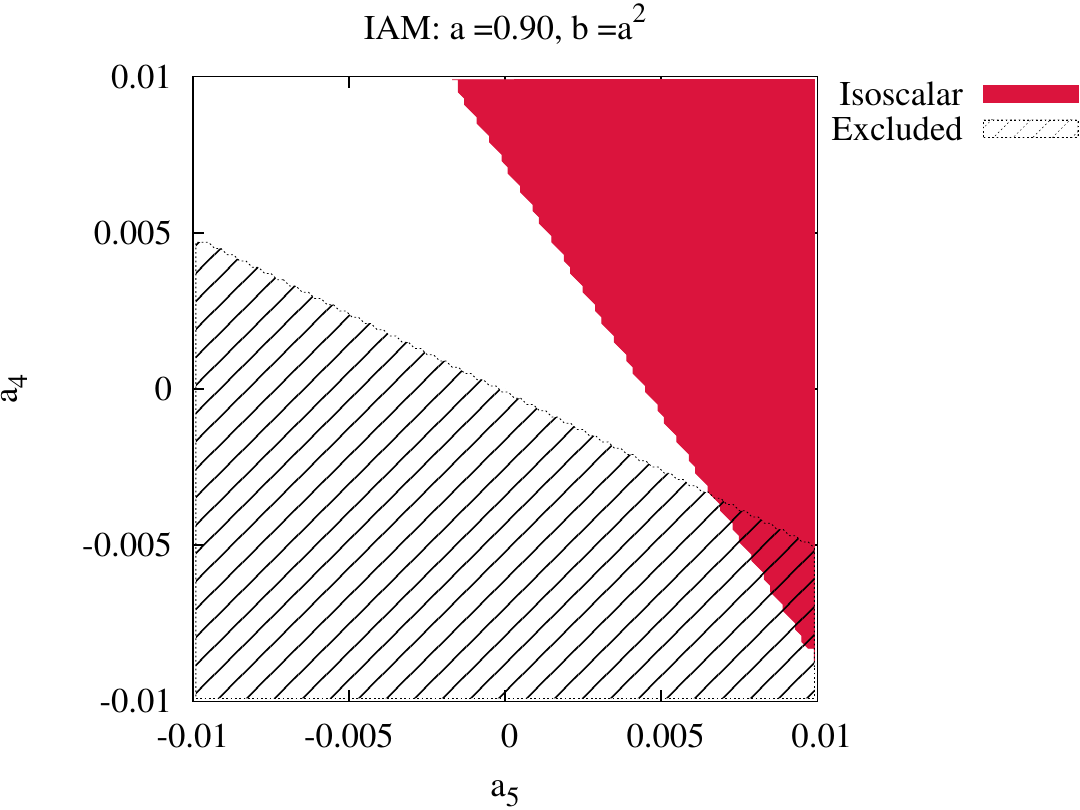}} \\
\caption{For $a=0.9$ and $b=a^2$:  (a) Regions with isoscalar and isovector resonances
 (and the isotensor \textit{exclusion} region) up to a scale $4\pi v \approx 3$~TeV.  (b) Same as (a), 
but only showing isoscalar/isovector resonances in which $M_{S,V}<600$~GeV. This can be compared with Fig. 
\ref{explot09} to conclude that $b$ has very little relevance here.\label{explot09ba2}}
\end{figure}

First of all we look for the existence of resonances. We set $b=1$ and consider two values~\footnote{
Other values of $a$ have also been studied but we here present results only for these two.} $a=0.9$ 
and $a=0.95$ compatible with experimental bound and indeed we easily find resonances in various channels. 
Most of them have the right 
causality properties that make the theory acceptable. However, in the $I=2,J=0$ channel we see 
that there is a region in the
$a_4-a_5$ plane where causality is violated. This corresponds to the shaded
region in the lower part of Figs. \ref{explot09},\ref{explot095} 
and \ref{explot09ba2} and the theories corresponding to these values for the parameters $a_4$, $a_5$
are therefore not acceptable. The presence of this excluded region is in exact 
correspondence with was found for the
$a=1$ case in \cite{Espriu:2012ih} (and also with a similar situation in pion physics\cite{iam}).

In Fig. \ref{explot09} we show the region of parameter space in $a_4$, $a_5$ 
where isoscalar and isovector resonances exist for the value $a=0.9$ 
along with the isotensor exclusion region. The pattern here has some analogies with the case
$a=1$ studied in \cite{Espriu:2012ih} but proper\footnote{Recall that resonances are required to 
have, in addition to the 
correct causal properties, $\Gamma < M/4$.} resonances are somewhat harder to form, in particular in 
the vector channel no resonance is found below 600 GeV for $a=0.9$ in contrast to
the $a=1$ case studied in \cite{Espriu:2012ih}. For $a=0.95$ some vector resonances can be found 
below 600 GeV for rather extreme values of $a_4$ and $a_5$ (upper left corner in Fig. \ref{explot095})

If no resonances are found at the LHC all the way up to 3 TeV, the values 
of $a_4$ and $a_5$ in the coloured regions in the left figures of Figs. \ref{explot09} and
 \ref{explot095} could be excluded and then $a_4$ and $a_5$ should lie within 
the small central region in the left plots of Figs. \ref{explot09} and \ref{explot095}.
Small as these regions are, they are noticeably larger than
the one corresponding to $a=1$, which was virtually non-existent. This is true even
for $a=0.95$ which is very close to the MSM value for $a$, $a=1$. 

We have also considered the case where $b=a^2$. In this case in the ET approximation the two channels decouple
and our implementation of the IAM becomes better justified. 
Results for the case $a=0.9$ and $b=a^2$ are depicted in Fig.
\ref{explot09ba2}. Changes with respect to $b=1$ are unnoticeable indicating that $b$ is 
of little relevance for the
presence of resonances.


\begin{figure}[tb]
\centering
\subfigure[(a)]{\includegraphics[clip,width=0.45\textwidth]{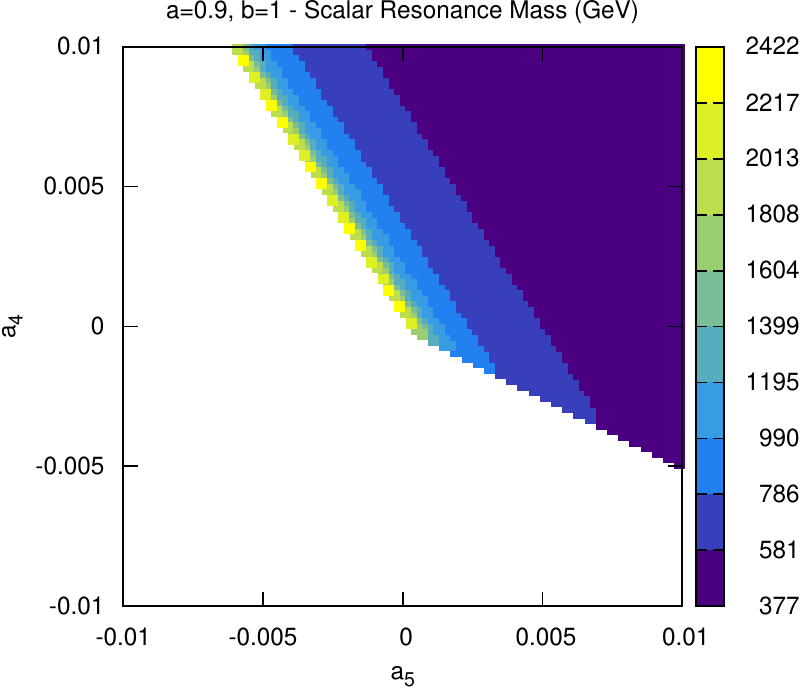}} \hspace{0.5cm}
\subfigure[(b)]{\includegraphics[clip,width=0.45\textwidth]{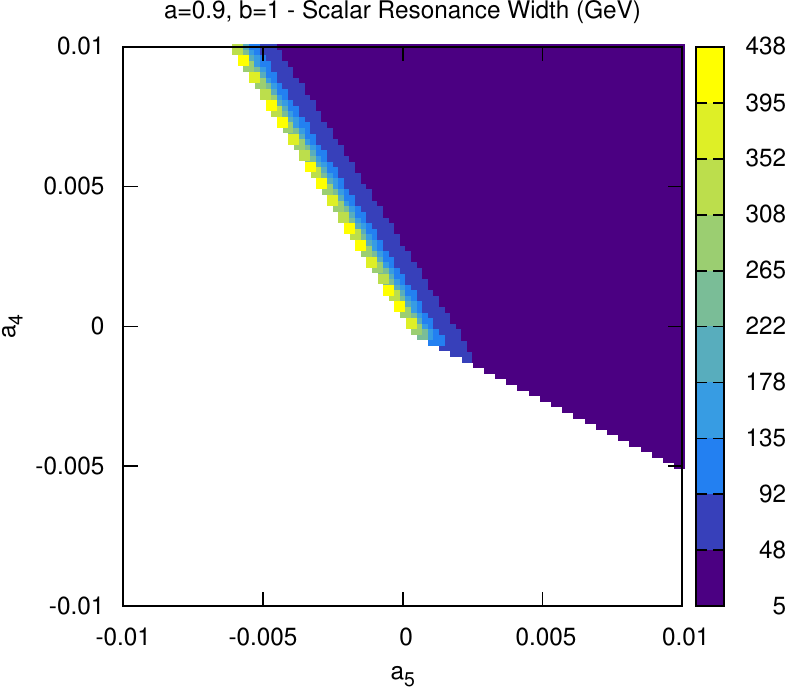}} \\
\subfigure[(c)]{\includegraphics[clip,width=0.45\textwidth]{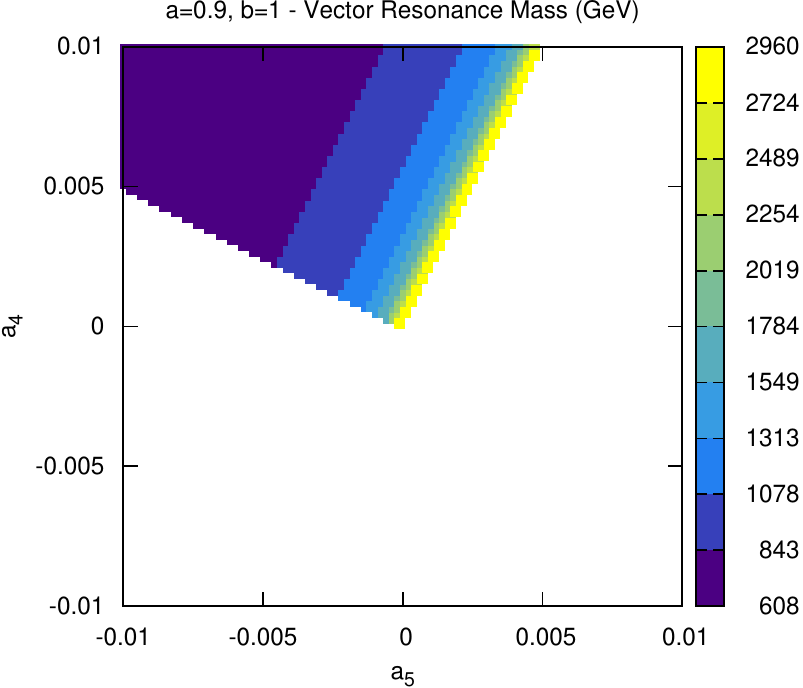}} \hspace{0.5cm}
\subfigure[(d)]{\includegraphics[clip,width=0.45\textwidth]{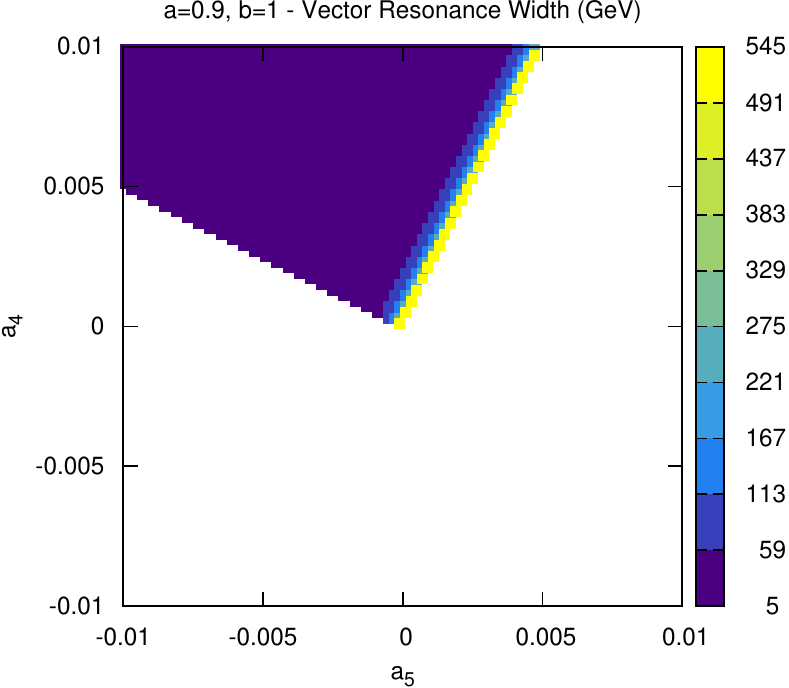}} \\
\caption{For $a=0.9$ and $b=1$, masses in GeV for (a) scalar and (b) vector resonances predicted from the unitarized 
partial wave amplitudes of $WW \to WW$ scattering.  Widths in GeV for the corresponding (c) scalar and (d) 
vector resonances.\label{contours09}}
\end{figure}

\begin{figure}[tb]
\centering
\subfigure[(a)]{\includegraphics[clip,width=0.45\textwidth]{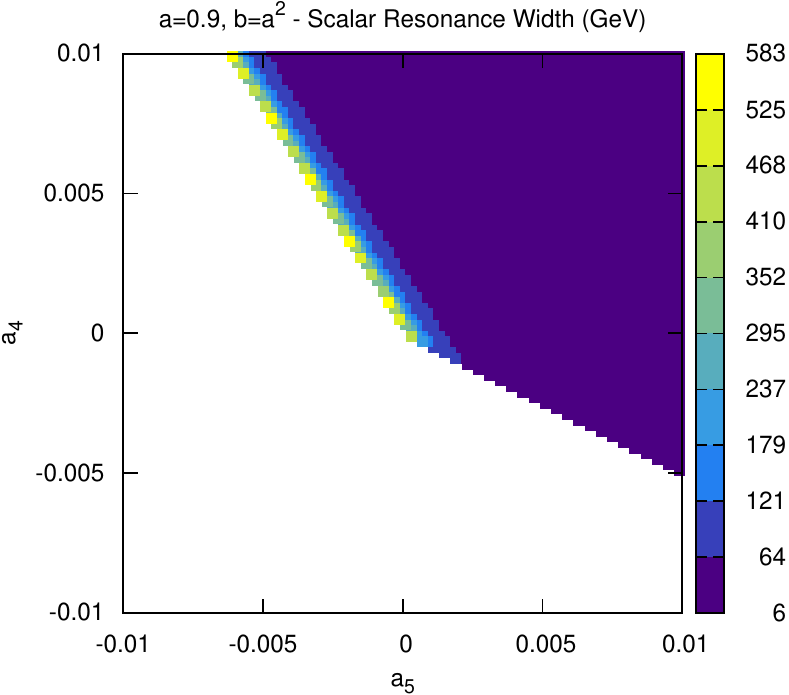}}
\subfigure[(b)]{\includegraphics[clip,width=0.45\textwidth]{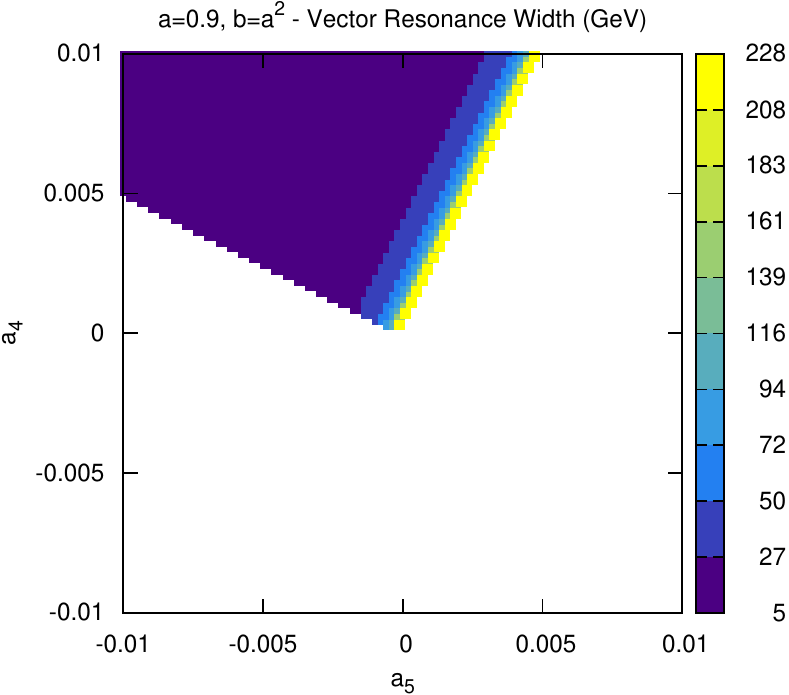}} 
\caption{For $a=0.9$ and $b=a^2$: Widths of the isoscalar (a) and isovector (b) resonances. Comparison 
with the equivalent plots of the previous figure shows some influence of $b$ here.\label{contours09ba2}}
\end{figure}

Figure \ref{contours09} shows the masses and widths of the scalar and vector resonances 
obtained for $a=0.9$. As we see, in general they
tend to be heavier and broader than the ones in the $a=1$ case studied in \cite{Espriu:2012ih}.
We emphasize that the resonance in the scalar channel is additional to the Higgs at 125 GeV.
The impact of parameter $b$ is actually more visible in the widths of the different resonances. In 
Fig. \ref{contours09ba2} we depict the widths obtained in the scalar and vector channels for $b=a^2$ when 
$a=0.9$ 

\subsection{Case $a>1$}
As we have seen the $a<1$ case is really a smooth continuation of the $a=1$ limit. Resonances are somewhat more
rare and they tend to be slightly heavier and broader, the more so as one departs from $a=1$ but
the modifications are small. This changes when we go to the $a>1$ case.

In this case the tree level amplitudes  $t_{IJ}^{(0)}$ tend to
$-\infty$ as $s\to \infty$. In the isoscalar channel for $1<a<1.125$ they possess several additional zeroes, 
which disappear for $a>1.125$. In the isovector channels the additional zeroes remain for even larger values
of $a$. Past these zeroes, 
the tree-level contribution is negative all the way up to the 
limit of validity of the effective theory. 

One finds zeroes of the denominator in eq.~(\ref{eq:t_iam}) that would correspond to
resonances provided that the numerator does not vanish. This comment is relevant because many of 
the resonances present, particularly in the vector channel, appear in  region near the last
(as $s$ increases) zero of the amplitude and this requires particular care. In fact for a set of values
of $a_4$ and $a_5$ the determination as to whether a pole exists or not becomes ambiguous.  

\begin{figure}[tb]
\centering
\subfigure[(a)]{\includegraphics[clip,width=0.45\textwidth]{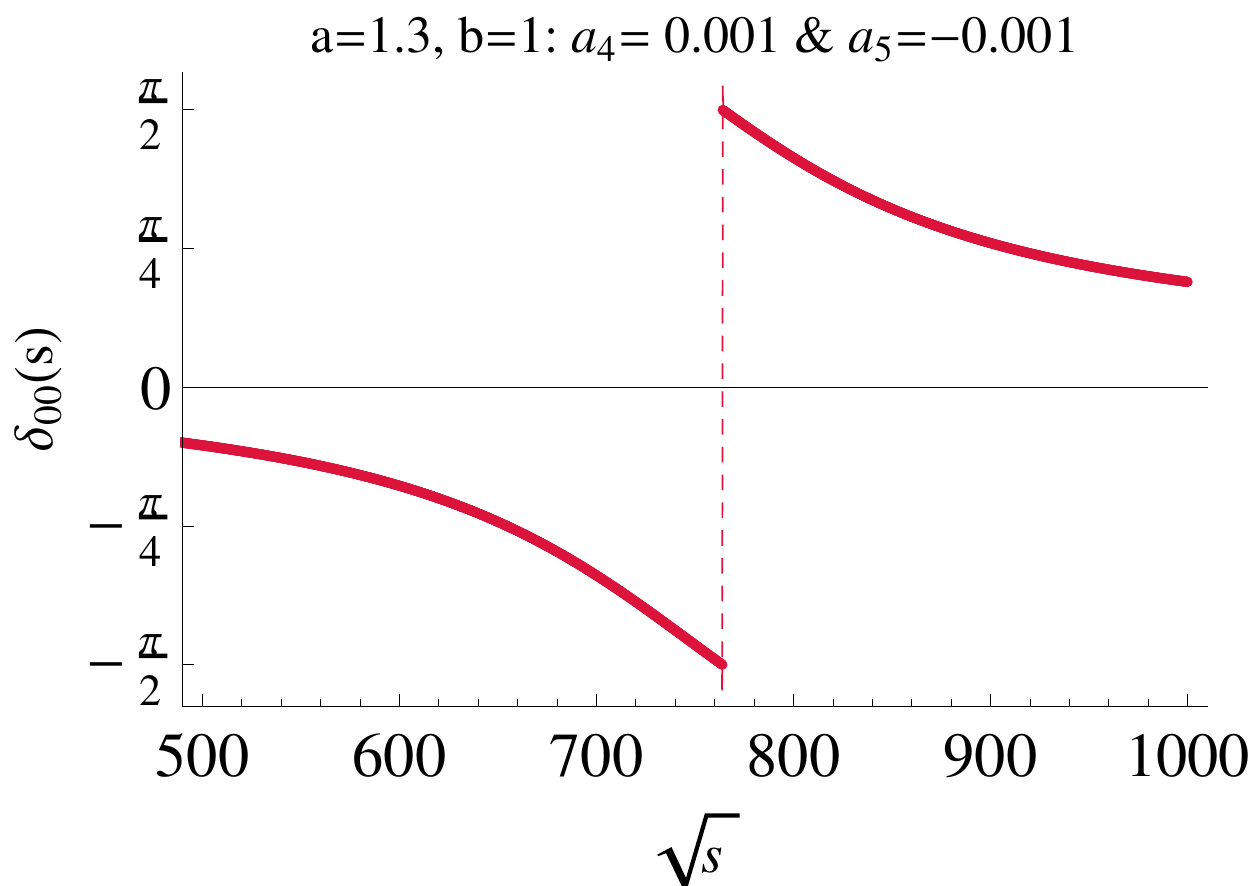}}
\subfigure[(b)]{\includegraphics[clip,width=0.45\textwidth]{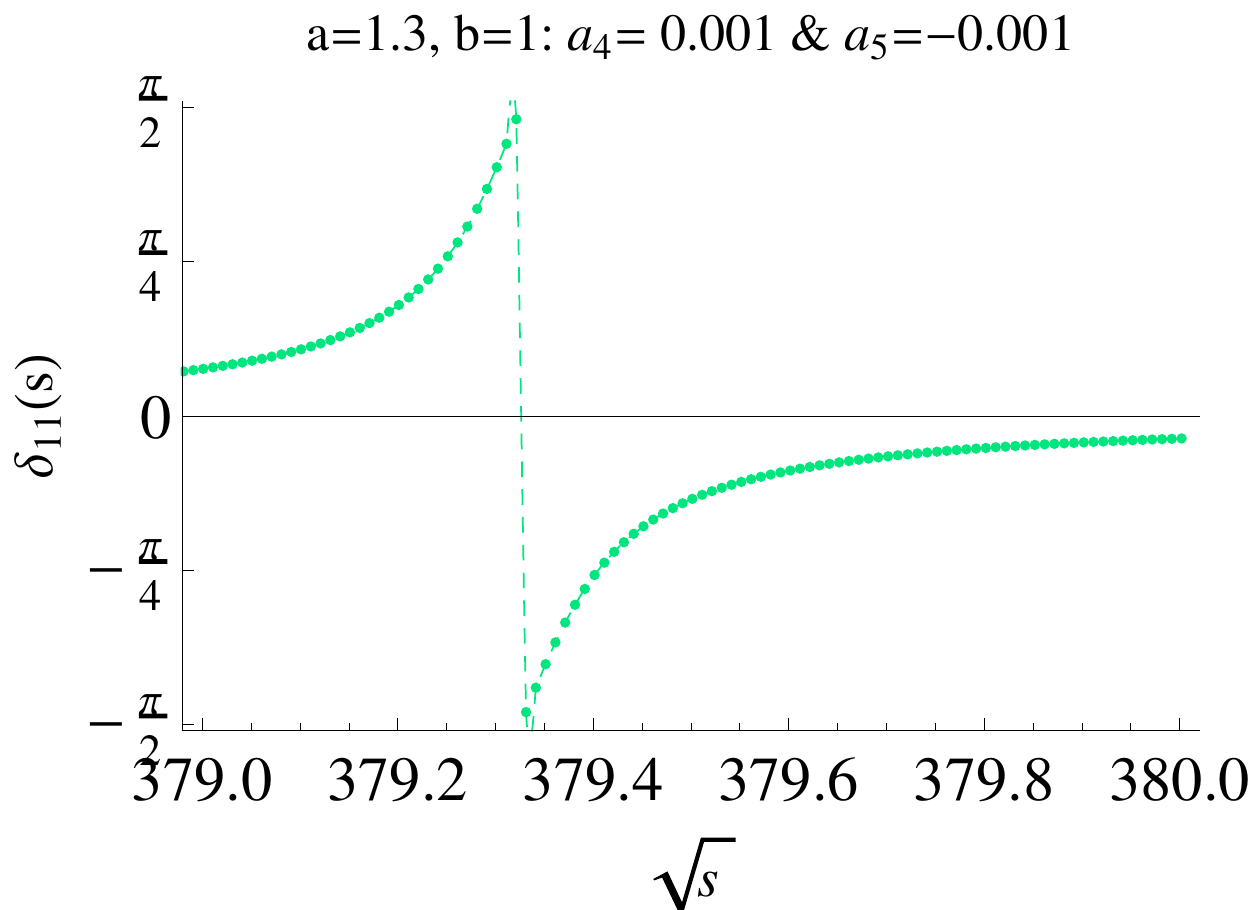}} 
\subfigure[(c)]{\includegraphics[clip,width=0.45\textwidth]{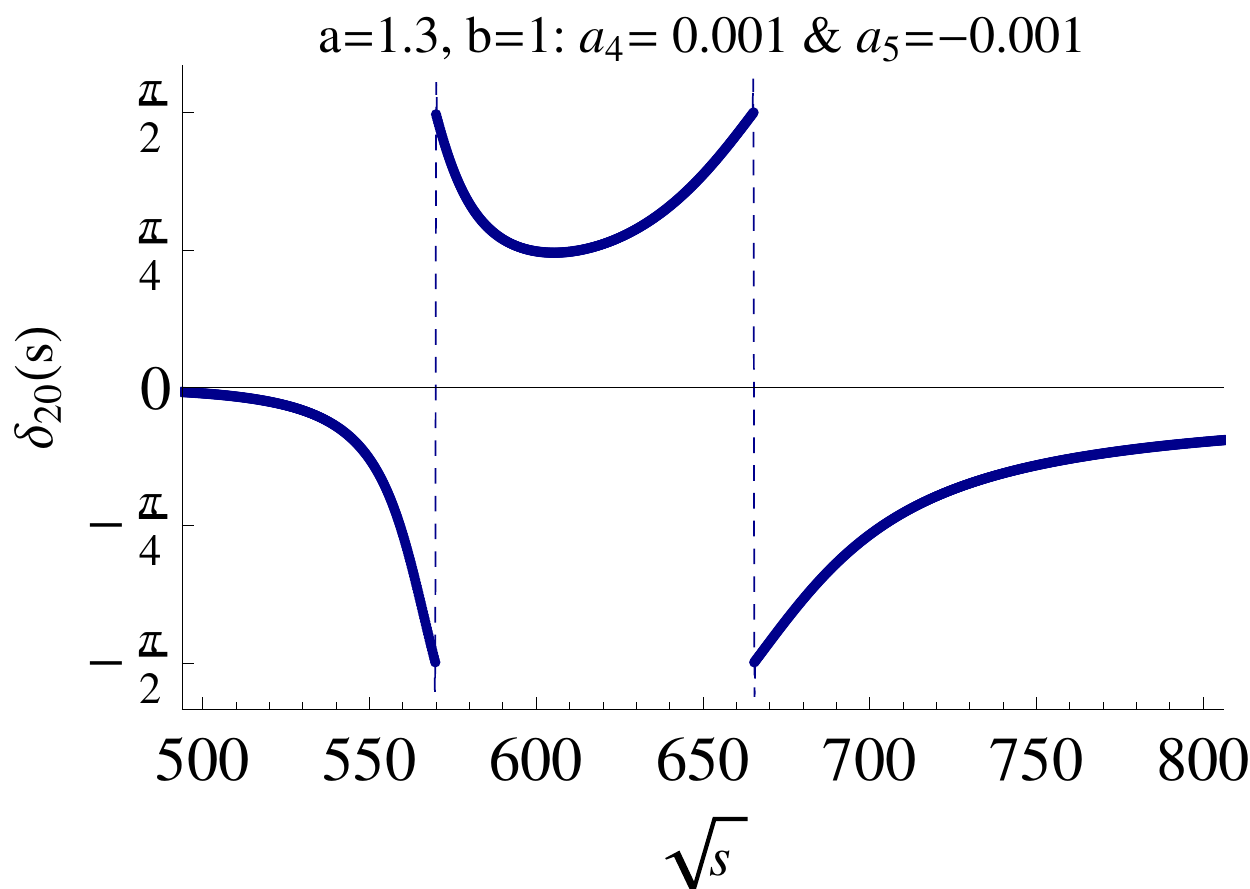}} 
\caption{Phase shifts for  $a_4= 0.001$ and $a_5= -0.001$ and
the values $a=1.3$ and $b=1$. The plots shows wrong resonances for isoscalar ($\simeq760$ GeV) and tensor ($\simeq570$ GeV) since 
the shift is from $-\pi/2$, whereas isovector has a good resoances ($\simeq380$ GeV). Moreover, the second tensor resonances 
($\simeq665$ GeV) with positive width is also shown.\label{pato}}
\end{figure}

When we continue our amplitudes into their second Riemann sheet to estimate
the width and solve for the complex pole we find that in various channels the imaginary part is 
such that it corresponds to a negative width. When two poles in a given channel are found, one is acceptable
but then the other one leads to acausal behaviour (this can be proven analytically). For other
values of the coupling the resonances are perfectly acceptable.
As an example of the pathologies found we show for $a=1.3$ in Figure~\ref{pato} the phase
shifts for isotensor-scalar and isovector channel. We can see a behaviour that
is incompatible with causality for the isotensor-scalar phase shift; recall that $\Gamma= 2 (\frac{d\delta}{d\sqrt{s}})^{-1}$.
\begin{figure}[tb]
\centering
\subfigure[(a)]{\includegraphics[clip,width=0.45\textwidth]{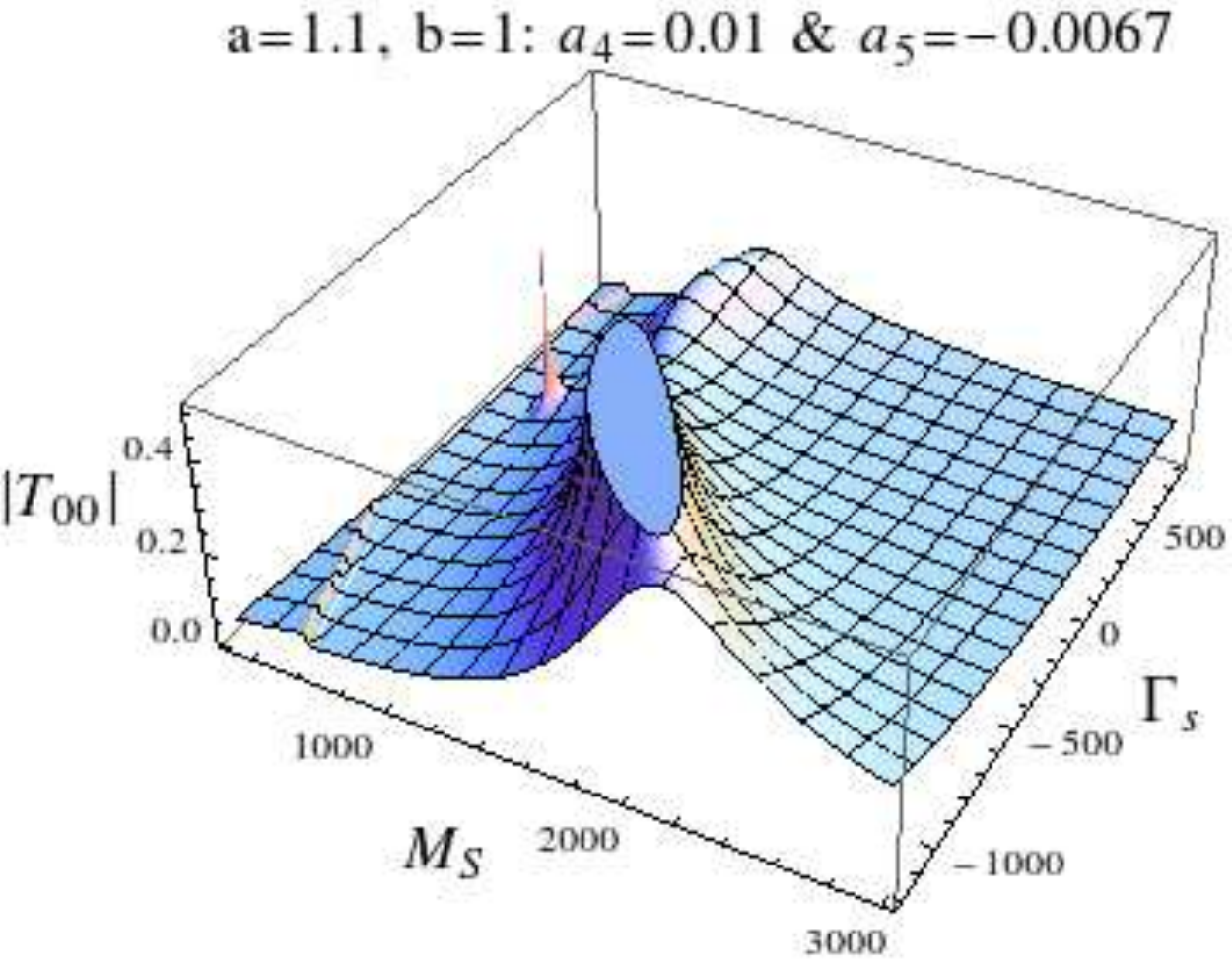}}
\subfigure[(b)]{\includegraphics[clip,width=0.45\textwidth]{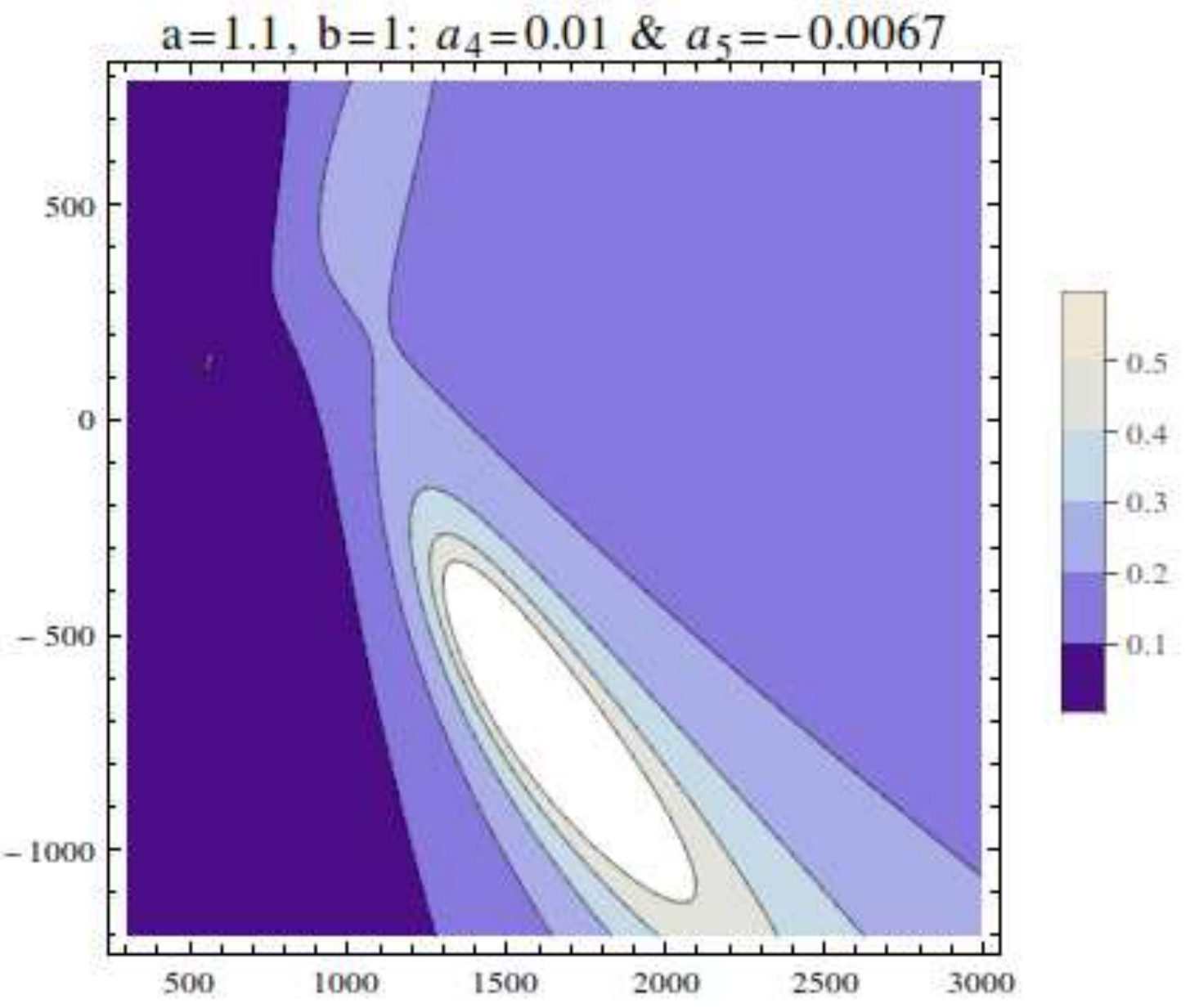}} 
\caption{(a) The two resonances  that appear in the scalar channel are shown in a 3D plot. The larger one
has a large negative width. The corresponding contour plot in shown in (b) where the physical one is nearly invisible
being extremely narrow.\label{diseases}}
\end{figure}
Sometimes a bona fide resonance pole coexists with a second resonance having negative width. This can be seen 
for instance in Fig.~\ref{diseases} in the scalar channel for $a=1.1$. We see that one genuine looking resonance
coexists with a huge singularity having a large negative width. The corresponding effective theory
is unacceptable. 
\begin{figure}[tb]
\centering
\subfigure[(a)]{\includegraphics[clip,width=0.45\textwidth]{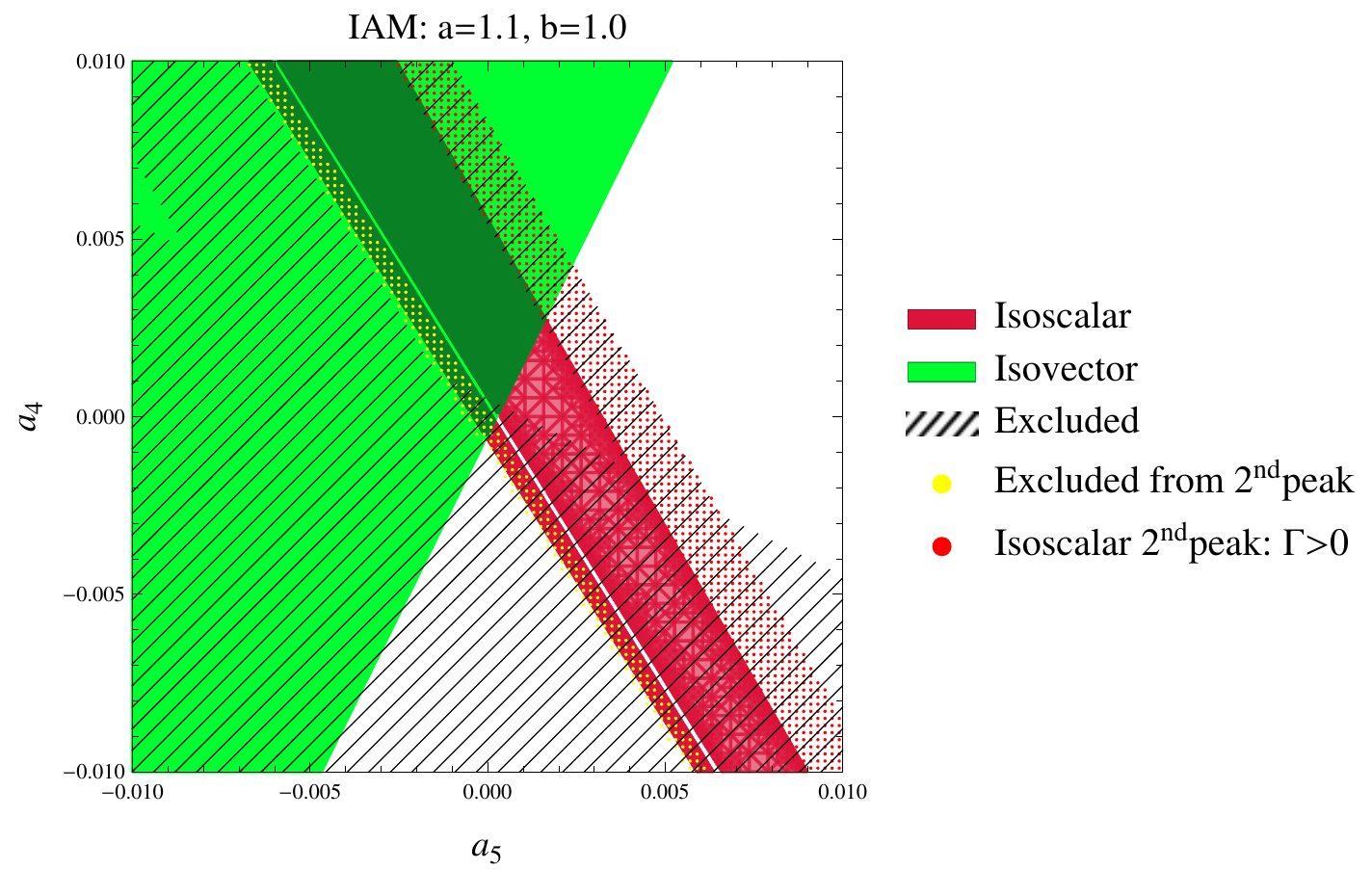}}
\subfigure[(b)]{\includegraphics[clip,width=0.45\textwidth]{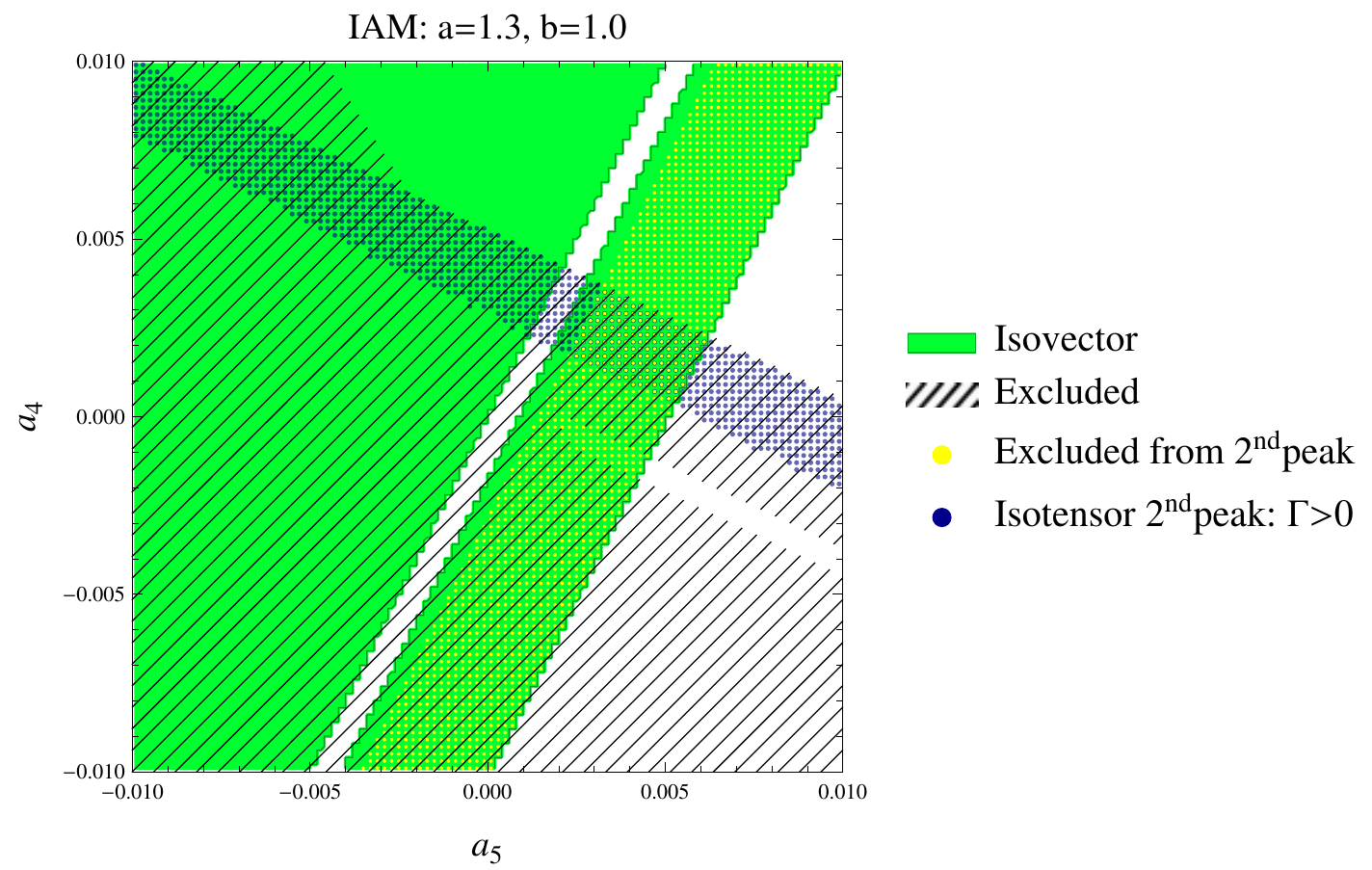}} 
\caption{(a) Search for 
resonances for $a=1.1$  up to the scale $4\pi v \approx 3$~TeV. The lower 
part of parameter space is excluded due to the isotensor resonance becoming acausal. In addition there is
an exclusion area due to unphysical poles in the $I=0$ channel. Some isovector and isoscalar resonances
are possible. In the white area in the upper left corner the scalar resonances are very broad and
are not considered as such by the $\Gamma< M/4$ condition. 
  (b) Same as (a) for $a=1.3$. The areas excluded due to resonances developing negative widths are now 
even larger. No resonance satisfying our criteria exists in the scalar channel, the apparent poles have all
negative widths for $I=0$. In a sizeable area vector resonances develop a second unphysical resonance. As for the
isotensor channel, most of the parameter space has one pole with negative width. Then a second exclusion
band (similar to the isovector one for $a=1.1$) exist due to isotensor channels having one valid resonance together
with a second acausal one. Only a small set of values present one valid resonance in the isovector channel.
Note that $a_4=a_5=0$ is unphysical for $a=1.3$. \label{explotagt1}}
\end{figure}

The net result is that a very sizeable part of the space of parameters is ruled out. For instance 
in Fig.~\ref{explotagt1} we show the excluded areas for $a=1.1$, Fig.~\ref{explotagt1}(a) and 
$a=1.3$, Fig.~\ref{explotagt1}(b). 
We are thus forced to conclude that pathologies abound in the $a>1$ case. In particular we have been unable to find a bona fide
$I=2$ dynamical resonance for $a=1.1$ and $a=1.3$ and this seems to be the generic situation for $a>1$. 

On the other hand, even though our findings  basically excludes $I=2$ dynamical resonances for $a>1$,
a light and elementary $I=2$ state can be included  in the lagrangian, such as in the Georgi-Machacek model with
a light quintuplet \cite{GM}. This is not contradictory to our findings.
If one wants to consider a weakly coupled
state with a mass much below the natural cut-off of the theory, it has to be made explicit in the effective
lagrangian as a propagating degree of freedom. Then it will appear as a pole also after unitarization, exactly
as the light Higgs does. In addtion there may be or not dynamical resonances depending on the nature
of the short distance theory. 
The IAM, that is known to work well for strongly coupled theories,
seems to be robust enough to support a strongly interacting sector and a perturbative 
one coexisting in the same theory. 

Interestingly, the difference betwen $a>1$ and $a<1$ regimes looks consistent to the 
sum rule introduced in~\cite{sumrule}. However, in the appendix we comment about a possible issue which 
could affect the derivation of this sum rule.

\section{Experimental visibility of the resonances}
One thing is having a resonance and a very different one is being able to detect it. In particular the
statistics so far available from the LHC experiments is limited. Searching for new particles in the LHC 
environment is extremely challenging. Yet a particle with the properties of the
Higgs has been found with only limited statistics.  This has been possible in part because of a fortunate upwards
statistical fluctuation but also because the couplings and other properties
of the Higgs were well known in the MSM. This is not necessarily the case for new resonances 
they may exist in the EWSBS. Fortunately the IAM method is able not only of predicting masses and widths but
also their couplings to the $W_LW_L$ and $Z_LZ_L$ channels. In \cite{Espriu:2012ih}, where the
case $a=1$ was considered, the experimental signal of the different resonances was compared to that of a MSM Higgs
with an identical mass. Because the decay modes are similar (in the vector boson channels that is) and limits on different 
Higgs masses are well studied this is a practical way of presenting the results. 

\begin{figure}[tb]
\centering
\subfigure[(a)]{\includegraphics[clip,width=0.4\textwidth]{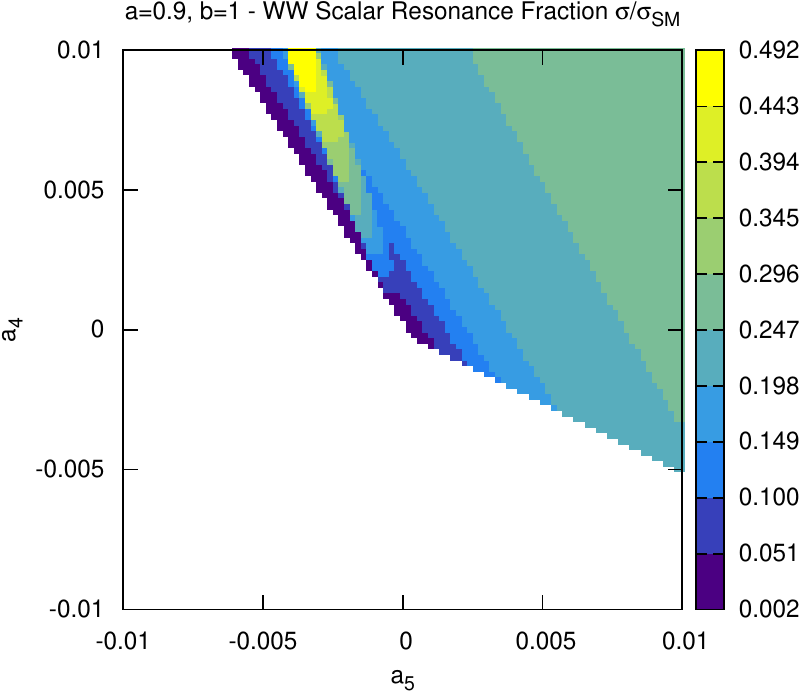}} 
\subfigure[(b)]{\includegraphics[clip,width=0.4\textwidth]{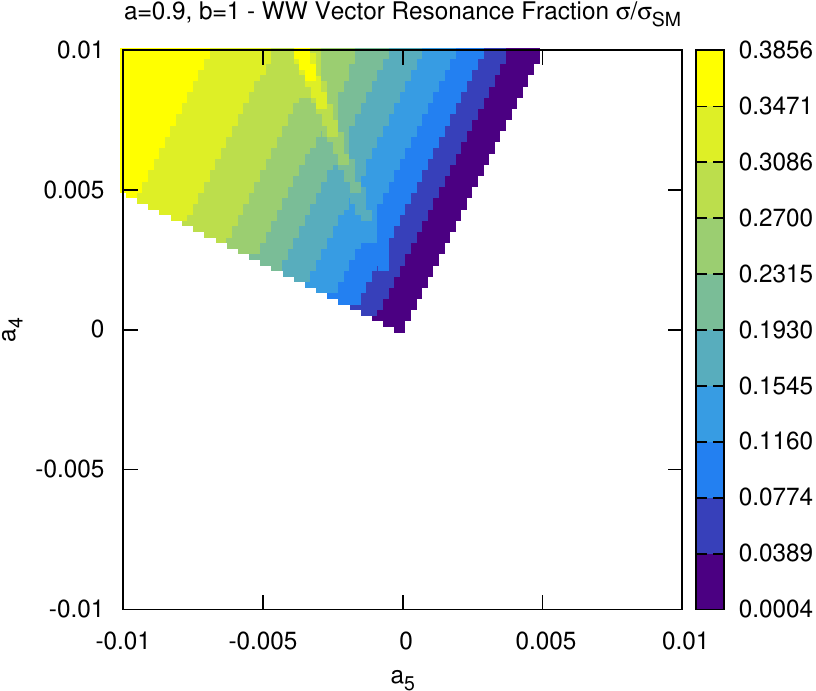}}
\subfigure[(c)]{\includegraphics[clip,width=0.4\textwidth]{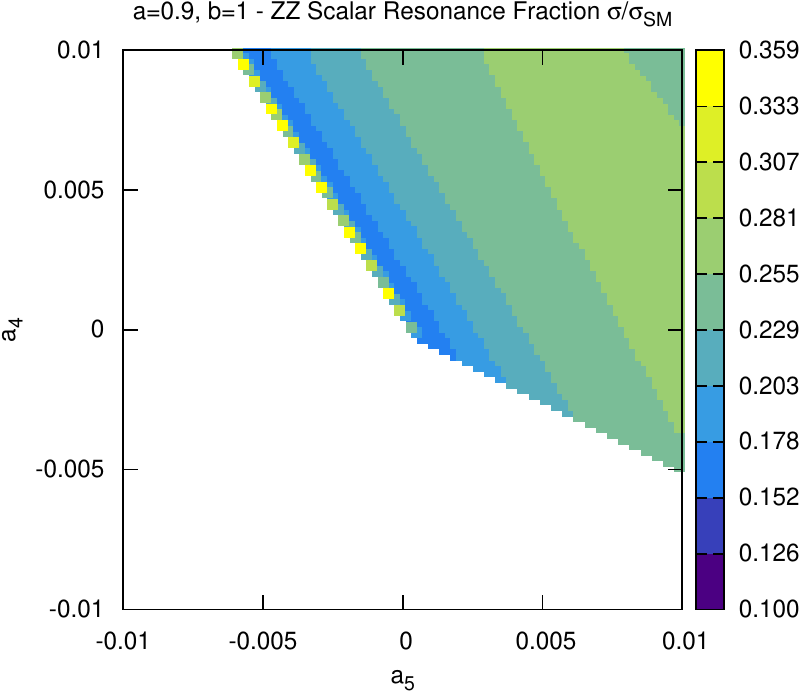}} 
\caption{For $a= 0.9$ and $b=1$:  Ratios of $W_LW_L$ scattering cross section due to dynamical 
resonances with that of the SM with a Higgs boson 
of the same mass for (a) scalar and (b) vector resonances, taken in the peak region as defined in
\cite{Espriu:2012ih}. Ratio of the $Z_LZ_L$ scattering cross section due to dynamical 
resonances with that of the SM with a Higgs boson 
of the same mass for a scalar resonance (c). The LHC energy has been taken to be 8 TeV and the EWA
approximation is assumed.\label{signal}}
\end{figure}

Therefore in order to gain some intuition as to whether any of the predicted resonances for $a<1$ should have
 been seen by now
at the LHC we compare their signal (the size of the corresponding Breit-Wigner resonance) with the  
one of the Higgs at an equivalent mass. Just to gain some intuition on this
we have used the easy-to-implement Effective W Approximation, or EWA \cite{ewa}. The
results are depicted in Fig. \ref{signal} for the $W_LW_L$ and 
$Z_L Z_L$ vector fusion channels.  Note that both production modes are sub-dominant at the LHC with respect to 
gluon production mediated by a top-quark loop and also note that the decay modes of the resonances can be predicted
with the technology presented here only for $W_LW_L$ and $Z_LZ_L$ final states.   

What can be seen in these figures is that the signal is always lower than the one for a Higgs boson of 
an equivalent mass. However, the ratio $\sigma_{\text{resonance}}/\sigma_{\text{Higgs}}$ seems to depend 
substantially on the value of $a$. For instance, for $a=1$ it was found that in the scalar channel this ratio
was typically lower than 0.1 and only in some very limited sector of parameter space could be as large
as 0.3. It was even lower for $ZZ$ production. Now for $a=0.9$ we see that 0.2 is a more typical value for
$\sigma_{\text{resonance}}/\sigma_{\text{Higgs}}$  and in some areas of parameter space can go up to $\sim 0.4$ or 
even close to 0.5. Again
the signal is somewhat lower in the $Z_LZ_L$ production channel.
For the vector channel and again normalizing to the Higgs signal we get ratios
 for $\sigma_{\text{resonance}}/\sigma_{\text{Higgs}}$ the signal ranges from 0.03 to 0.3.

\section{Conclusions}
In this paper we have extended the analysis of \cite{Espriu:2012ih} to the case $a\neq1$ and $b\neq 1$  imposing
the require of unitarity on the fixed isospin amplitudes contributing to longitudinal $W$ scattering. The method
chosen to unitarize the partial waves is the Inverse Amplitude Method. The simplicity of this method makes it
suitable to analyze the problem being considered, while its validity has been  well tested in strong
interactions in the past. 

We have seen that even in the presence of a light Higgs, it can help constrain anomalous couplings by 
helping predict heavier resonances, present in an extended EWSBS. The results for $a\neq 1$ presented here
turn out to be partly in line with the results for $a=1$ previously obtained if $a<1$
and partly qualitatively different if $a>1$ . If $a<1$ 
for a large subset of values of the higher dimensional
coefficients resonances are present. Typically they tend to be heavier and broader than in the $a=1$ case.   
but only moderately so. They are never like the broad resonances that were entertained in the past in Higgsless
models and this is undoubtedly a consequence of the unitarization that the presence of the Higgs brings about. There
is a smaller room for new states once unitarity is required.
The properties of the resonance are therefore radically different from the initial expectations concerning $W_LW_L$ scattering

Current LHC Higgs search results do not yet probe the IAM resonances, but it may be possible in the near future, this is
particularly true if $a$ departs from its Standard Model value $a=1$ because the resonances become higher and broader
in this case with values for the ratio to the experimental signal that a Higgs with an equivalent mass would give
$\sigma_{\text{resonance}}/\sigma_{\text{Higgs}}$ can get close to 0.5 (recall that this applies only to the longitudinal vector
gauge boson fusion channel). In any case it seems that LHC@14 TeV will be able to probe a reasonable part of the 
possible parameter space for resonances.  

If resonances are found with the properties predicted here this discovery  would immediately 
indicate that there is an extended EWSBS and that
 this is likely described by some strongly interacting theory, giving credit to the hypothesis of the Higgs being
a composite state ---most likely a pseudo-Goldstone boson. It would also provide immediate information on the value
of some of the higher dimensional coefficients in the effective theory, probably much earlier that direct $W_LW_L$ scattering
would allow for a determination of the quartic gauge boson coupling.

We have also found another interesting result, namely that in the present framework theories with $a>1$ are nearly
excluded as the IAM predicts that they lead to resonances that violate causality in a large part of parameter space, the more
so as one departs more from $a=1$.

\section*{Acknowledgements}
We thank A. Dobado, J. Gonz\'alez-Fraile, M.J. Herrero, J.R. Pel\'aez, B. Yencho  for discussions concerning different aspects of
unitarization and effective lagrangians. 
We acknowledge the financial support from projects FPA2010-20807, 2009SGR502 and CPAN (Consolider CSD2007-00042).



\appendix

\section{Tree-level  $W_LW_L $ scattering amplitudes} 
\label{sec:appendix_amplitudes}
\noindent 
In the isospin limit, $M=M_{Z} = M_{W}$, and with massive $W$, the tree-level and $a_{4,5}$-dependent 
amplitude for $\wplus_{L} \wminus_{L} \to Z_{L} Z_{L}$ scattering is given by 
\bea\label{fullamp}
A_{W^{+}W^{-} \to ZZ}^{ {\rm tree} \, + \, a_{i}} \left( p_{1},p_{2},p_{3},p_{4} \right)
& = & 
\;
-2 g^{2}(1  - g^{2}a_{5} ) 
\eeee{1}{2}{3}{4} \\ \nn
&& +  g^{2}(1 + g^{2}a_{4}) \Big[
\eeee{1}{4}{2}{3} + \eeee{1}{3}{2}{4} 
\Big]
\\ \nn
& &
+g^{2} \Bigg\{ \left(\frac{1}{(p_{1}-p_{3})^{2}-M^{2}}\right)
\Bigg[ \\ \nn
& & 
-4 \Big( 
\eepepe{1}{2}{1}{3}{2}{4} + \eepepe{1}{4}{1}{3}{4}{2} + \\ \nn
& & \hspace{0.75cm} 
\eepepe{2}{3}{3}{1}{2}{4} + \eepepe{3}{4}{3}{1}{4}{2}
\Big) \\ \nn
& & 
+2\Big(
\edote{2}{4} \Big(
\pdote{1}{3}(p_{2}+p_{4})\cdot\epsilon_{1} +
\pdote{3}{1}(p_{2}+p_{4})\cdot\epsilon_{3}
\Big) + \\ \nn
& & \hspace{0.75cm} 
\edote{1}{3} \Big(
\pdote{2}{4}(p_{1}+p_{3})\cdot\epsilon_{2} +
\pdote{4}{2}(p_{1}+p_{3})\cdot\epsilon_{4}
\Big) \Big)  \\ \nn
& & \hspace{0.45cm}
- \eeee{1}{3}{2}{4} \Big(
(p_{1}+p_{3})\cdot p_{2} + (p_{2}+p_{4})\cdot p_{1}
\Big)
\Bigg] \\ \nonumber
& &
+ \hspace{0.5cm} (p_{3} \leftrightarrow p_{4})
\Bigg\} 
-a^2 g^{2} M^{2}\left(\frac{\edote{1}{2} \edote{3}{4}}{(p_{1}+p_{2})^{2}-M_{H}^{2}} \right) \, , \nn
\eea
where $\epsilon_{i} = \epsilon_{L}(p_{i})$. The analogous expression in the ET approximation is much simpler
\be\label{etamp}
A_{w^{+} w^{-} \to z z}^{ {\rm tree } \, + \, a_{i}} (s)
 = - \fracp{s}{v^{2}} \left(\frac{(a^{2}-1)s + M_{H}^{2}}{s-M_{H}^{2}}
 -2 \fracp{s}{v^{2}}\left(a_4 (1+\cos^2\theta) + 4 a_5\right) \right)
 \ee
For completeness, we also give the amplitude for the $\wplus_{L} \wminus_{L} \to h h$ scattering 
\bea
\!\!A_{W^{+}W^{-} \to hh}^{{\rm tree }}(p_1,p_1,q_3,q_4) &=& 
g^2 \left(\frac{b}{2} \edote{1}{2} - \frac{3 a M_H^2}{2(M_H^2-(p_1+p_2)^2)}\edote{1}{2}\right. \\
\no
&+&\!\!\!\!\left.a^2  \frac{g^2 v^2}{4 M^2}\left(\frac{\edotqmp{1}{3}{1}\edotqmp{2}{3}{1}
 -M^2 \edote{1}{2}}{M^2-(q_3-p_1)^2} + (q_3\leftrightarrow q_4)\right)\right),
\eea
In the CM reference frame the expression for 
$A_{W^+W^-\to ZZ}^{\rm{tree}\, + \, a_{i}}(s,t,u)$ becomes 
\bea\label{eq:CMS-A}
A_{W^+W^-\to ZZ}^{\rm{tree}\, + \, a_{i}}(s,t,u) &=& 
\frac{a^2 \left(s-2 M^2\right)^2}{v^2 \left(M_H^2-s\right)}\\
&+&\frac{768
   M^{10}-128 M^8 (5 s+4 t)+32 M^6 \left(7 s^2+8 s t+4 t^2\right)}
{v^2 \left(s-4 M^2\right)^2 \left(M^2-t\right) \left(-3M^2+s+t\right)}\nn\\
&-& \frac{8 M^4 s 
   \left(5 s^2+11 s t+4 t^2\right)-  
   M^2 s^2 \left(3 s^2+18 s t+14 t^2\right)+s^3 t
   (s+t)}
   {v^2 \left(s-4 M^2\right)^2 \left(M^2-t\right) \left(-3 M^2+s+t\right)}\nn  \\ 
      &+&\frac{8 a_5 (s-2 M^2)^2 + 2 a_4 (16 M^4 - 8 M^2 s + (1 + \cos^2\theta) s^2))}{v^4}\nn
   \eea
Recall that $A_{W^+W^-\to ZZ}^{\rm{tree} \, + \, a_{i}}(s,t,u)$
for the scattering of longitudinally polarized $W$ is not Lorentz invariant. The expression
above is valid in CM frame only.

\section{The issue of crossing symmetry}
\label{sec:appendix_crossing}

We would like to clarify the issue of crossing symmetry of amplitudes 
with external $W_L$'s. To this end let us consider
just the tree-level contribution in the MSM to the processes $W_L^+ W_L^-\to W_L^+ W_L^-$ and
$W_L^+ W_L^+\to W_L^+ W_L^+$, respectively. 

To keep the formulae simple while making the point let us consider the
limit $s\to \infty$, $-t\to \infty$ in the first process, which is consistent except for $\cos\theta \simeq 1$,
and expand in $M^2/s$ and $M^2/t$. We borrow the results from \cite{esma}. 
The resulting amplitude is
\be
-g^2\left(\frac{M_H^2}{4M^2}\left[\frac{t}{t-M_H^2}+\frac{s}{s-M_H^2}\right]
+\frac{s^2+t^2+st}{2st}
-\frac{M_H^2}{s}\frac{2M_H^2t-s(s+t)}{(M_H^2-s)(M_H^2-t)}\right)+ \ldots.
\ee
In the second process we expand in powers of $M^2/u$ and $M^2/t$. One then gets
\be
-g^2\left(\frac{M_H^2}{4M^2}\left[\frac{t}{t-M_H^2}+\frac{u}{u-M_H^2}\right]
+\frac{u^2+t^2+ut}{2ut}
+\frac{M_H^2}{t+u}\frac{(t-u)^2}{(M_H^2-u)(M_H^2-t)}\right)+ \ldots.
\ee
The two processes are related by crossing and one would naively think that the two amplitudes can be 
related by simply exchanging $s$ and $u$. While this is correct for the first two terms in both equations, 
it fails for the third.
If the reader is worried about the approximations made, more lengthy complete results are given in \cite{esma} and
they show the same features.

The reason is that while crossing certainly holds when exchanging the external four vectors, the reference frame
in which the two above amplitudes are expressed are different. In both cases they correspond to center-of-mass 
amplitudes, but after the exchange of momenta the two systems are boosted one with respect
to the other. Writing the amplitudes in terms of $s,t,u$ gives
the false impresion that these expressions hold in any reference system but this is not correct
because the polarization vectors are no true four-vectors.

On the contrary, the amplitudes computed via the ET are manifestly crossing symmetric because they amount
to replacing $\epsilon_L^\mu \to k^\mu$, which is obviously a covariant 4-vector. We insist once more that crossing
{\em does hold} in any case but is not {\em manifest} for the scattering of longitudinal $W$ bosons at the
level of Mandelstam variables.

\section{The origin of the logarithmic poles} 
\label{sec:appendix_logarithmic}

Here we discuss the origins of the $3$ singularities at $s_0=M^2_H$, $s_1=4 M^2 - M_H^2$ and  $s_2=3 M^2$) 
entering the $t_{IJ}(s)$ amplitudes. These singularities  can be tracked back from the terms 
$1/(s-M^2_H)$, $1/(t-M^2)$ and $1/(u-M^2)$ in the $W^+_LW^-_L\to Z_L Z_L$ amplitude in eq.~\ref{eq:CMS-A}. 
The origin of the pole at $s_0$  is fairly obvious and needs no justification.
 
As for the other two singularities, the term $1/(t-M^2)=1/((-1 + \cos\theta) (-4 M^2 + s)/2 -M^2)$ has 
a pole at $s_3$ for $\cos\theta=-1$ which under integration in $\cos\theta$ to derive the partial
wave amplitude $t_{IJ}(s)$ becomes a logarithmic pole as well as for $1/(u-M^2)=1/((1 + \cos\theta) (-4 M^2 + s)/2 -M^2)$ 
at $\cos\theta=-1$.  This explains the presence of $s_2$ pole for $t_{IJ}(s)$ amplitudes.

The origin of the pole at $s_1$ for  $t_{IJ}(s)$ amplitudes is more complicated to see. First of all, let us notice that
the fixed-isospin amplitudes  $T_I$ in eq.~\ref{eq:fixed_isospin} are combinations of 
the $A^{++00}=A(W^+_LW^-_L\to Z_L Z_L)$ in eq.~\ref{eq:CMS-A} and its crossed amplitude  $A^{++++}=A(W^+_LW^+_L\to W^+_LW^+_L)$. 
At this point, the term $1/(s-M^2_H)$  in $A^{++00}$, eq.~\ref{eq:CMS-A}, 
trasforms for the crossed amplitude $A^{++++}$ into  $1/(t-M^2_H)= (1/(-1 + \cos\theta) (-4 M^2 + s)/2 -M_H^2)$. 
Then, for $\cos\theta=-1$ we have a pole at $s_1$ and under integration on $\cos\theta$ the amplitude $t_{IJ}(s)$
gets a logarithmic pole at $s_1$.

Note that these singularities are all below threshold. Note too that except
for $s_0$ they are absent in the ET treatment. For the LHC they appear at values of $s$ corresponding to
the replacement $4 M^2 \to \sum_i q_i^2$ as the external $W$ are typically off-shell.

\section{Sum rule}
\label{sec:appendix_sumrule}
In \cite{sumrule} the following sum rule was derived
\be\label{sum}
\frac{1-a^2}{v^2} = \frac{1}{6\pi}\int_0^\infty \frac{ds}{s}
\left(2\sigma_{I=0}(s)^{tot} + 3 \sigma_{I=1}(s)^{tot} -5
\sigma_{I=2}(s)^{tot}\right) + c_\infty,
\ee
where $\sigma_I^{tot}$ is the total cross section in the isospin channel $I$ and $c_\infty$
is the contribution of the $\vert s\vert \to \infty$ contour to the dispersive integral. This latter
contribution can sometimes be neglected. This is the case for instance in $\pi\pi$ scattering. In the forward
direction it is expected to show a Regge behaviour compatible with the neglection of the external part
of the circuit. 

The interesting result (\ref{sum})
was derived making full use of the Equivalence Theorem and setting $M=0$. 
As we have seen, at low $s$ there are some relevant deviations with respect to the ET predictions when using
the proper longitudinal vector boson amplitudes and including the $t$-channel $W$ exchange, and they affect 
the analytic properties of the amplitude.
Let us see how this sum rule is affected by these deviations.

The technique used in \cite{sumrule} to derive the previous result was to define 
the function $F(s,t,u)=A_{W^+W^-\to ZZ}^{\rm{tree}}(s,t,u)/s^2$, consider the
case $t=0$, corrresponding to the forward amplitude, and compute the integral 
\be
\oint ds F(s,t,u)
\ee
using two different circuits: one around the origin and another one along the cuts in the real axis and
closing at infinity (this last contribution actually drops if the amplitudes are assumed to grow slower than
$s$). 

Applying the strict ET, each order in perturbation theory contributes to a given order in an expansion
in powers of $s,t,u$. Therefore if the integral is done
in a small circle around the origin only the tree-level amplitude eq.~(\ref{etamp}) contributes and 
taking both contributions into account results in the
result on the left hand side of  equ.~(\ref{sum}). On the other hand, the integral along the left cut
can be related using crossing symmetry to the one on the right cut and eventually leads to the right
hand side of equation~(\ref{sum}).

Formulae (\ref{fullamp}) and (\ref{etamp}) show clearly that the analytic structure of the full result and the ET 
one are quite different at low values of $s$. In the exact case and for the tree-level amplitude 
we have four poles for $F(s,t,u)$. We assume that $s$ and $t$ are independent variables
and to make this visible we replace $t\to \bar t$ 
\bea
\!\!\!\!\!\!\!\!\!\!\!\!\!\!\!\!\!\!s_0 = 0 &\to& \text{Res}(F(s_0,\bar t,u))=
\frac{4 a^2 M^2 (M^2-M_H^2)}{M_H^4
   v^2}+\frac{2 \bar t \left(8 M^4-7 M^2\bar  t+\bar t^2\right)}{v^2
   \left(M^2-\bar t\right) \left(\bar t-3 M^2\right)^2}\nn\\
s_1 = M_H^2 &\to& \text{Res}(F(s_1,\bar t,u))=
-\frac{a^2 \left(M_H^2-2 M^2\right)^2}{M_H^4 v^2}\nn\\
s_2 = 3 M^2 - \bar t 
&\to& \text{Res}(F(s_2,\bar t,u))=
-\frac{-27 M^8+52 M^6 \bar t+M^4 \bar t^2+2 M^2\bar  t^3}{v^2 \left(\bar t-3
   M^2\right)^2 \left(M^2+\bar t\right)^2}\nn\\
   s_3 = 4 M^2 &\to& \text{Res}(F(s_3,\bar t,u))=
   \bar t\frac{10 M^4 - 3 M^2 \bar t - 3 \bar t^2}{(M^2 - \bar t) (M^2 + \bar t)^2 v^2}\\
&& \sum_{i=0,3} \text{Res}(F(s_i,\bar t,u)) =
 \frac{(3 - a^2) M^2 - (1 - a^2)\bar  t}{(M^2 - \bar t) v^2} \label{B2}
 \eea
Note however that for $s=s_3=4 M^2$, the $t$ variable is always zero, being 
$t= -(1-\cos\theta)(s - 4 M^2)/2$, and $u= -(1+\cos\theta)(s - 4 M^2)/2$.  
This shows that $s$ and $t$ are dependent for some exceptional kinematical points, 
for example when the inicial states are at rest ($s=s_3=4 M^2$).
Therefore when $s\to s_3$,  $t\to 0$. If we set $\bar t=0$ at the outset the sum of residues leads to
\be\label{residuesfinal}
\sum_{i=0,3} \text{Res}(F(s_i,\bar t=0,u)) =
 \frac{(3 - a^2)}{v^2}\,.
\ee
which differs from the result quoted in \cite{sumrule}. The reason is clear when looking at eq.~(\ref{B2}): if we take 
the limit $M\to 0$ at the outset as is done in the strict ET approximation, we get one
result, while if $ \bar t$ is set to zero with $M\neq 0$, we get a different one.

In addition, in a complete calculation (as opposed to the simpler ET treatment) it is not true that a 
given order in the chiral expansion corresponds
to a definite power of $s$. Therefore,  when $M$ is not neglected the order $s$ contribution will have corrections
from all orders in perturbation theory. The contribution to the left hand side of the integral, obtained after
circumnavigating all the poles will then be of the form 
\be 
 \frac{3 - a^2 + \mathcal{O}(g^2)}{v^2}\,.
\ee
Actually, the right cut changes too when $M$ is taken to be non-zero; it starts at $s=4M^2$ (which is not a pole
as we have just discussed because it has a vanishing residue). The left cut is not changed as for $t=0$ the
$u$ channel has a cut for $s<0$ corresponding to $u> 4M^2$. 

Although crossing symmetry is not manifest 
(see appendix~\ref{sec:appendix_crossing}) for the full amplitudes it remains valid\footnote{We thank the 
referee
for pointing this out to us.} for $t=0$ 
and it is possible to relate exactly the contribution along the left cut to
the analogous integral along the right one.


\end{document}